\documentclass[aps,twocolumn,superscriptaddress,showpacs,showkeys,preprintnumbers,amsmath,amssymb,amsfonts,nofootinbib]{revtex4}

\pdfoutput=1

\usepackage{graphicx}
\usepackage{color}
\usepackage{slashed}
\usepackage{tabularx}
\usepackage{hyperref}
\usepackage{bbold}
\usepackage{bbm}
\usepackage{amsmath}
\usepackage{placeins}
\usepackage[normalem]{ulem}

\newcommand{\tr}{\mathrm{tr}}
\newcommand{\Tr}{\mathrm{Tr}}
\newcommand{\STr}{\mathrm{STr}}

\newcommand{\I}{\mathrm{i}}
\newcommand{\Nf}{N_{\text{f}}}
\newcommand{\Nc}{N_{\text{c}}}
\newcommand{\sgn}{\operatorname{sgn}}
\renewcommand{\Re}{{\text{Re}\,}}
\renewcommand{\Im}{{\text{Im}\,}}

\newcolumntype{L}[1]{>{\raggedright\arraybackslash}p{#1}} 
\newcolumntype{C}[1]{>{\centering\arraybackslash}p{#1}} 
\newcolumntype{R}[1]{>{\raggedleft\arraybackslash}p{#1}} 

\graphicspath{{.}{figures/}{mma/}}

\begin{document}

\title{Fermionic spectral functions with the Functional Renormalization Group}

\newcommand{\ECT}{European Centre for Theoretical Studies in Nuclear Physics and related Areas (ECT*) and Fondazione Bruno Kessler, Villa Tambosi, Strada delle Tabarelle 286, I-38123 Villazzano (TN), Italy}
\newcommand{\JLU}{Institut f\"ur Theoretische Physik, Justus-Liebig-Universit\"at Giessen, Heinrich-Buff-Ring 16, 35392 Giessen, Germany}
\newcommand{\TUDa}{Institut f\"ur Kernphysik (Theoriezentrum), Technische Universit\"at Darmstadt, Schlossgartenstr. 2, 64289 Darmstadt, Germany}

\author{Ralf-Arno Tripolt}\affiliation{\ECT}
\author{Johannes Weyrich}\affiliation{\JLU}
\author{Lorenz von Smekal}\affiliation{\JLU}
\author{Jochen Wambach}\affiliation{\ECT}\affiliation{\TUDa}

\begin{abstract}
We present first results on the calculation of fermionic spectral functions from analytically continued flow equations within the Functional Renormalization Group approach. Our method is based on the same analytic continuation from imaginary to real frequencies that was developed and used previously for bosonic spectral functions. In order to demonstrate the applicability of the method also for fermionic correlations we apply it here to the real-time quark propagator in the quark-meson model and calculate the corresponding quark spectral functions in the vacuum.
\end{abstract}



\maketitle

\section{Introduction}\label{sec:introduction}

The spectral properties of strongly interacting matter under extreme conditions, as encountered in the early universe and compact stellar objects, are of fundamental importance for identifying the relevant degrees of freedom in the equation of state and respective transport properties. Spectral functions are real-time quantities, while the underlying equilibrium state is commonly obtained from imaginary-time (Euclidean) evaluations of the partition function. In this setting, a thermodynamically consistent computation of the spectral properties poses a major challenge since analytic continuations of the pertinent Euclidean $n$-point functions are required. In relativistic theories this entails a transition from Euclidean to Minkowski space-time, see for example \cite{Vidberg:1977,Jarrell:1996,Asakawa:2000tr,Dudal2013}.

In the context of the strong interaction (QCD) quark spectral functions are of particular interest. A Bayesian reconstruction method has, for example, been used in \cite{Qin:2013ufa} and \cite{FischerPawlowskiRothkopfEtAl2017} to extract quark spectral functions from Euclidean data obtained from Dyson-Schwinger equations. Within the Functional Renormalization Group (FRG), which incorporates thermal as well as quantum fluctuations, Euclidean quark propagators have recently been calculated in \cite{FischerPawlowskiRothkopfEtAl2017} and \cite{CyrolMitterPawlowskiEtAl2017}.
In the present work, we focus on the calculation of real-time quark propagators. Instead of using numerical reconstruction methods, see e.g.~\cite{TripoltGublerUlybysheEtAl2018, Cyrol:2018xeq}, we perform the analytic continuation on the level of the FRG flow equations for retarded two-point correlation functions which are then solved directly in the corresponding domain of frequencies close to the real axis. Such analytic continuation methods have been put forward in \cite{Floerchinger2012} and \cite{Strodthoff:2011tz, Kamikado2013,Kamikado2014}. In \cite{Tripolt2014, Tripolt2014a, TripoltSmekalWambach2017} the approach was extended to finite temperature, finite quark chemical potential as well as to finite spatial momenta and has been used to calculate mesonic spectral functions within the quark-meson model. The analytically continued FRG flow equations for the corresponding two-point correlation functions were thereby solved in a simple but thermodynamically consistent and symmetry preserving truncation  which in the long-wavelength and static limit reduces to the leading-order derivative expansion used for the underlying effective potential. In \cite{JungRenneckeTripoltEtAl2017} this approach was extended to calculate in-medium vector- and axial-vector meson spectral functions. For the first time we here present an FRG calculation of fermionic spectral functions obtained from analytically continued flow equations which can be solved numerically.
In a first step we restrict ourselves to the vacuum and to vanishing external spatial momenta.

This work is organized as follows. In Sec.~\ref{sec:FRG_QM} we briefly introduce the FRG framework and its application to the quark-meson model. Our analytic continuation method and the flow equation for the real-time quark propagator and the spectral functions are discussed in Sec.~\ref{sec:analytic_continuation}. We have solved the flow equations using both, a grid and a Taylor-expansion method as discussed in Sec.~\ref{sec:numeric_implementation} where we also demonstrate the particular advantages and disadvantages of either method. Results for the quark mass and the mass dressing function are presented in Sec.~\ref{sec:results1} while respective results for the quark propagator and the quark spectral functions are shown in Sec.~\ref{sec:results2}. Various sum rules, which can be derived from the Lehmann representation of the quark propagator, are discussed in Sec.~\ref{sec:sumrules}. We close with our summary and outlook in Sec.~\ref{sec:summary}. Further details are collected in an appendix.

\section{Functional renormalization group and quark-meson model}\label{sec:FRG_QM}

The Functional Renormalization Group (FRG) is a non-perturbative approach that is used for example in quantum and statistical field theory, in particular for strongly interacting systems, see e.g.~\cite{Berges:2000ew,Polonyi:2001se,Pawlowski:2005xe,Schaefer:2006sr,Kopietz2010,Braun:2011pp, Friman:2011zz, Gies2012} for reviews. It is formulated in (continuous) Euclidean space-time and combines Wilson's idea of the renormalization group in momentum space \cite{Wilson1971, Wilson1974} with functional methods in quantum field theory.

In the following we will use the formulation pioneered by Wetterich \cite{Wetterich:1992yh} which aims at calculating the effective average action $\Gamma_k$ where $k$ is the renormalization group scale. At the ultraviolet (UV) scale $k=\Lambda$, the effective average action is basically given by the bare action $S$ of the chosen model and does not include any fluctuation effects. By lowering the scale $k$ the effects of quantum and thermal fluctuations are gradually included until the full effective action $\Gamma=\Gamma_{k=0}$ is obtained in the limit $k\rightarrow0$.
The scale-dependence of $\Gamma_k$ is given by the following flow equation, also known as the Wetterich equation,
\begin{align}
\label{eq:Wetterich}
\partial_k \Gamma_k[\phi,\psi,\bar{\psi}]=
&\frac{1}{2}\,\STr \left[\partial_kR_k \left(\Gamma_{k}^{(2)}[\phi,\psi,\bar{\psi}]+R_k\right)^{-1}\right],
\end{align}
where $R_k$ is a regulator function that suppresses momentum modes with momenta smaller than $k$,\footnote{While the FRG flow for the effective average action explicitly contains the regulator $R_k$, physics at $k\to 0$ should not depend on a particular choice. For an up-to-date discussion of how to devise optimized regulators in a particular truncation where this can be quite non-trivial, see \cite{Pawlowski:2015mlf}.} and $\Gamma_{k}^{(2)}$ is the second functional derivative with respect to the fields. Both $\Gamma_{k}^{(2)}$ and $R_k$ can be represented as matrices in the field space of bosonic and fermionic variables, and the supertrace runs over field space as well as all internal indices also including an integration over internal momenta. 

We will apply this flow equation to the quark-meson model as a low-energy effective theory for the chiral aspects of QCD with two flavors \cite{Jungnickel:1995fp,Schaefer:2004en}. It includes quarks, the sigma meson and the pions as effective degrees of freedom which interact via a Yukawa-type interaction. We use the following ansatz for the effective average action of the quark-meson model in Euclidean space-time,
\begin{align}\label{eq:gamma}
\Gamma_{k}[\phi, \psi, \bar\psi]=
\int d^{4}x \:\Big\{
&\bar{\psi}\left(\gamma_\mu\partial^\mu+
h(\sigma+i\vec{\tau}\vec{\pi}\gamma^{5}) \right)\psi\nonumber\\
&+\frac{1}{2} (\partial_{\mu}\phi)^{2}+U_{k}(\phi^2)-c\sigma
\Big\},
\end{align}
with $\phi^2=\sigma^2+\vec\pi^2$. This approximation, which is the leading order in a derivative expansion where the only scale-dependent object is the effective potential $U_{k}(\phi^2)$, is also called the local potential approximation (LPA) \cite{Litim2001,Braun:2009si}. When inserting this ansatz into the Wetterich equation, one obtains the flow equation for the effective potential,
\begin{align}
\label{eq:flow_pot}
\partial_k U_k =\frac{1}{2} I_{k,\sigma}^{(1)} +\frac{3}{2} I_{k,\pi}^{(1)} -\Nc \Nf I_{k,\psi}^{(1)},
\end{align}
where explicit expressions for the threshold functions $I_k$ are given in App.~\ref{app:defs}. At the UV scale $\Lambda$ we choose the effective potential to be symmetric,
\begin{equation}
\label{eq:pot_UV}
U_\Lambda(\phi^{2}) =
\frac{1}{2}m_\Lambda^{2}\phi^{2} +
\frac{1}{4}\lambda_\Lambda(\phi^{2})^{2},
\end{equation}
and then solve the corresponding flow equation numerically, see Sec.~\ref{sec:numeric_implementation}. The term $c\sigma$ which breaks chiral symmetry explicitly and thus plays the role of the (up/down) current quark mass in QCD, is added to the effective potential in the infrared (IR) while spontaneous chiral symmetry breaking occurs dynamically through the fermionic fluctuations which are included by solving the flow equation. The solution for the scale-dependent effective potential is then used as input for the calculation of the fermionic two-point function.

\begin{figure*}[t!]
	\includegraphics[width=0.9\textwidth]{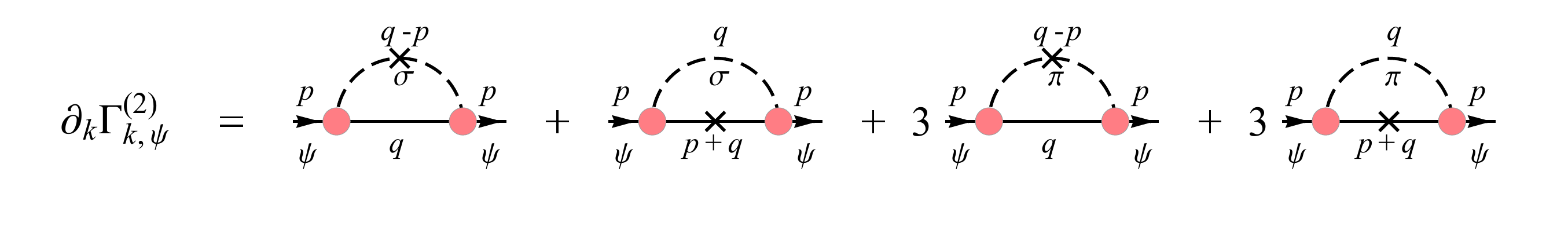}\vspace*{-5mm}
	\caption{Diagrammatic representation of the flow equation for the quark two-point function, Eq.~(\ref{eq:flow_gamma2}). Solid lines represent quark propagators, dashed lines meson propagators. The crosses represent regulator insertions $\partial_kR_k$ and the red circles the appropriate vertex functions.}
	\label{fig:flow_Gamma2}
\end{figure*}

In order to obtain the flow equation for the quark two-point function we take two functional derivatives of the Wetterich equation, Eq.~(\ref{eq:Wetterich}), with respect to the fermionic fields which gives
\begin{align}
\label{eq:flow_gamma2}
&\partial_k\Gamma_{k,\bar\psi\psi}^{(2)}=\frac{1}{2}\Tr\Big(\nonumber\\
&\:\:\:\:\:\:\partial_k R_B(\vec{q}-\vec{p})D_{\phi\phi}(q-p)\Gamma^{(3)}_{\bar\psi \psi \phi}D_{\bar\psi \psi}(q)\Gamma^{(3)}_{\bar\psi \psi \phi}D_{\phi \phi}(q-p)\nonumber\\
&\:+\partial_k R_F(\vec{q}+\vec{p})D_{\bar\psi \psi}(q+p)\Gamma^{(3)}_{\bar\psi \psi \phi}D_{\phi\phi}(q)\Gamma^{(3)}_{\bar\psi \psi \phi}D_{\bar\psi \psi}(q+p)\nonumber\\
&\:+\partial_k R_B(\vec{q}-\vec{p})D_{\phi\phi}(q-p)\Gamma^{(3)}_{\bar\psi \psi \phi}D_{\bar\psi \psi}(q)\Gamma^{(3)}_{\bar\psi \psi \phi}D_{\phi \phi}(q-p)\nonumber\\
&\:+\partial_k R_F(\vec{q}+\vec{p})D_{\bar\psi \psi}(q+p)\Gamma^{(3)}_{\bar\psi \psi \phi}D_{\phi\phi}(q)\Gamma^{(3)}_{\bar\psi \psi \phi}D_{\bar\psi \psi}(q+p)\Big),
\end{align}
see Fig.~\ref{fig:flow_Gamma2} for a diagrammatic representation. Therein, $q=(q_0,\vec{q})$ is the internal and $p=(p_0,\vec{p})$ the external momentum, $D=(\Gamma_{k}^{(2)}+R_k)^{-1}$ is the full regulated propagator, the vertex functions $\Gamma^{(3)}_{\bar\psi \psi \phi}$ are obtained from the ansatz in Eq.~(\ref{eq:gamma}), and the remaining trace represents a summation over all internal indices as well as an integration over internal momenta. Explicit expressions for the vertex functions $\Gamma^{(3)}_{\bar\psi \psi \phi}$ as well as for the bosonic and fermionic regulator functions, $R_B$ and $R_F$, are given in App.~\ref{app:defs}. As in the original studies \cite{Kamikado2014,Tripolt2014,Tripolt2014a}, here we use three-dimensional regulator functions which only regulate spatial momenta but not the energy components at the expense of some breaking of the Euclidean $O(4)$ symmetry. This breaking was assessed and found to be negligible for external momenta well below the UV cutoff scale $\Lambda$ in the Euclidean two-point functions, with still reasonably small and only quantitative effects in the time-like domain after analytic continuation \cite{Kamikado2014}. While this can be avoided in principle \cite{Pawlowski:2015mia,Pawlowski:2017gxj}, the three-dimensional regulators allow to perform the integration over the internal energy component or the corresponding Matsubara sum at finite temperature analytically which tremendously simplifies the analytic continuation procedure discussed in the next section.

\section{Analytic continuation and spectral functions}\label{sec:analytic_continuation}

In the UV, the Euclidean quark two-point function is given by
\begin{align}
\Gamma^{(2),E}_{\Lambda,\psi}(p_0,\vec{p})= -i\gamma_0p_0 -i\vec{\gamma}\vec{p}+h\sigma  \, ,\label{EuclideanGamma2}
\end{align}
with hermitian $\gamma$-matrices as can also be seen from Eq.~(\ref{eq:gamma}). We perform the analytic continuation by treating the Euclidean energy variable as a general complex argument $z = ip_0$, so that with $z=\omega+i \epsilon$ one obtains 2-point functions with retarded boundary conditions in the limit $\epsilon\rightarrow 0$.\footnote{In the imaginary parts of the flow equations for two-point functions the limit $\epsilon\to 0$ can be taken exactly, see App.~\ref{app:defs}.} We note that the resulting retarded propagator is analytic in the upper-half complex energy plane, as expected. To reproduce the standard form of the Dirac operator we furthermore introduce an overall minus sign as compared to Euclidean conventions adopted in (\ref{EuclideanGamma2}), to write
\begin{align}
\label{eq:continuation}
\Gamma^{(2)}_{\Lambda,\psi}(z,\vec{p})=  \gamma_0 z + i\vec{\gamma}\vec{p} - h\sigma \, ,
\end{align}
where we formally kept the Euclidean $\vec\gamma$'s which are usually changed to be anti-hermitian (by replacing $\vec\gamma\to i\vec\gamma$) in Minkowski space-time, of course. Based on this Dirac structure, we therefore make the following ansatz for the scale-dependent quark two-point function,
\begin{align}
\label{eq:Gamma2_R}
\Gamma^{(2)}_{k,\psi}(\omega,\vec{p})=\gamma_0C_k(\omega,\vec{p}) + i\vec{\gamma}\hat{\vec p}\:A_k(\omega,\vec{p})- B_k(\omega,\vec{p})\, ,
\end{align}
with $\hat{\vec p}\equiv\vec{p}/|\vec{p}|$. The dressing functions can be obtained from the full two-point function as follows,
\begin{align}
A_{k}(\omega,\vec{p}) &= - \frac{1}{4} \, \tr \Big( i\vec{\gamma}\hat{\vec p}\: \Gamma^{(2)}_{k,\psi}(\omega,\vec{p})\Big)\, ,\\
B_{k}(\omega,\vec{p}) &= - \frac{1}{4} \, \tr \Big( \Gamma^{(2)}_{k,\psi}(\omega,\vec{p})\Big)\, ,\\
C_{k}(\omega,\vec{p}) &= \frac{1}{4} \, \tr \Big( \gamma_0 \Gamma^{(2)}_{k,\psi}(\omega,\vec{p})\Big)\, .
\end{align}
The UV initial conditions for the dressing functions are thus given by
\begin{align}
\label{eq_UV_values_coeffs}
A_{\Lambda}(\omega,\vec{p}) &= |\vec{p}|\,,\\
B_{\Lambda}(\omega,\vec{p}) &= h\sigma\, , \label{initialB} \\
C_{\Lambda}(\omega,\vec{p}) &= \omega \, . \label{initialC}
\end{align}
The flow equation for the quark two-point function,
\begin{align}
  \partial_k\Gamma^{(2)}_{k,\psi}(\omega,\vec{p})=&\gamma_0\partial_kC_k(\omega,\vec{p}) + i\vec{\gamma}\hat{\vec{p}}\:\partial_kA_k(\omega,\vec{p})\nonumber\\
  &\quad -\partial_kB_k(\omega,\vec{p})\, ,
\end{align}
then leads to flow equations for the individual  dressing functions
of the form
\begin{align}
\label{eq:flow_eq_2PF}
\partial_kX_k(\omega,\vec{p})&=
\mathcal{J}^{(X)}_{k,\sigma\psi}(\omega,\vec{p})+
\mathcal{J}^{(X)}_{k,\psi\sigma}(\omega,\vec{p})\nonumber\\
&\quad \: \: \: +
3\,\mathcal{J}^{(X)}_{k,\pi\psi}(\omega,\vec{p})+
3\,\mathcal{J}^{(X)}_{k,\psi\pi}(\omega,\vec{p})\, ,
\end{align}
with $X\in \{A,B,C\}$. The explicit expressions for the generalized loop functions $\mathcal{J}_k$ herein are given in App.~\ref{app:defs}. In particular, the analyticity of the flow of these dressing functions in the upper-half of the complex energy plane is evident from these expressions, cf.~Eqs.~(\ref{eq:JB})-(\ref{eq:JC2}), and guarantees that the correct analytic behavior of the retarded propagator is maintained in the flow.

For later convenience we also write down corresponding expressions for the inverse of the quark two-point function in Eq.~(\ref{eq:Gamma2_R}), because we will need the imaginary parts of the retarded quark propagator for the various spectral functions,
\begin{align}
G_{k,\psi}(\omega,\vec{p})=& \gamma_0G^{(C)}_{k,\psi}(\omega,\vec{p})
+i\vec{\gamma}\hat{\vec p}\:G^{(A)}_{k,\psi}(\omega,\vec{p}) 
+ G^{(B)}_{k,\psi}(\omega,\vec{p})\, , \nonumber
\end{align}
with
\begin{align}
G^{(A)}_{k,\psi}(\omega,\vec{p})&=\frac{A_k(\omega,\vec{p})}{C_k^2(\omega,\vec{p})-A_k^2(\omega,\vec{p})-B_k^2(\omega,\vec{p})}\,,\\
G^{(B)}_{k,\psi}(\omega,\vec{p})&=\frac{B_k(\omega,\vec{p})}{C_k^2(\omega,\vec{p})-A_k^2(\omega,\vec{p})-B_k^2(\omega,\vec{p})}\, ,\\
G^{(C)}_{k,\psi}(\omega,\vec{p})&=\frac{C_k(\omega,\vec{p})}{C_k^2(\omega,\vec{p})-A_k^2(\omega,\vec{p})-B_k^2(\omega,\vec{p})}\, .
\end{align}
Note that this is not the regularized propagator $D=(\Gamma_{k}^{(2)}+R_k)^{-1}$ used in the loops, but the (retarded) inverse of $\Gamma_{k,\psi}^{(2)} $ in  (\ref{eq:Gamma2_R}). As such, in the UV, it is given by
\begin{align}
G_{\Lambda,\psi}(\omega,\vec{p})=
\frac{\gamma_0\omega+i\vec{\gamma}\vec{p} + h\sigma}
{(\omega+i\epsilon)^2-\vec{p}^2 - h^2\sigma^2}\, ,
\end{align}
as usual. At zero temperature, one generally has
\begin{equation}
  A_k(\omega,\vec p) = |\vec p| Z_k(p^2)\, ,\;\; C_k(\omega,\vec p) = \omega Z_k(p^2)   \, ,
  \label{eq:Z_k}
\end{equation}
i.e., both are determined by a single dimensionless renormalization function $Z$ of the invariant four-momentum $p^2$. $G^{(A)}_{k,\psi}(\omega,\vec{p})$ and $G^{(C)}_{k,\psi}(\omega,\vec{p})$ are then essentially the same, likewise. Here we keep them formally distinct, nevertheless, so that their flow equations can be readily extended to finite temperature and density in the future.

The quark spectral function can be obtained from the retarded propagator by taking the imaginary part,
\begin{align}
\rho_{k,\psi}(\omega,\vec{p})=-\frac{1}{\pi}\Im G_{k,\psi}(\omega,\vec{p})\,,
\end{align}
and therefore has the same Dirac structure as the propagator and the 2-point function,
\begin{align}
\rho_{k,\psi}(\omega,\vec{p})=\gamma_0\rho^{(C)}_{k,\psi}(\omega,\vec{p})+i\vec{\gamma}\hat{\vec{p}}\:\rho^{(A)}_{k,\psi}(\omega,\vec{p})+\rho^{(B)}_{k,\psi}(\omega,\vec{p})\, .
\end{align}
In the UV, the quark spectral function is given by
\begin{align}
\rho_{\Lambda,\psi}(\omega,\vec{p})=
\sgn (\omega)(\gamma_0\omega +i\vec{\gamma}\vec{p}+ h\sigma)\delta(\omega^2-\vec{p}^2-h^2\sigma^2)\, . \label{initialSpec}
\end{align}
The individual components of the spectral function can then be obtained directly from the imaginary parts of the corresponding propagator components,
\begin{align}
\rho^{(X)}_{k,\psi}(\omega,\vec{p})&=- \frac{1}{\pi}\Im G^{(X)}_{k,\psi\bar{\psi}}(\omega,\vec{p})\, ,
\end{align}
with $X\in \{A,B,C\}$. At zero temperature, $\rho^{(A)}_{k,\psi}(\omega,\vec{p})$ and $\rho^{(C)}_{k,\psi}(\omega,\vec{p})$ are essentially the same as well, one then usually writes,
\begin{equation}
\rho^{(C)}_{k,\psi}(\omega,\vec{p}) = \omega  \rho^{(Z)}_{k,\psi}(p^2)\, ,\;
\;  \rho^{(A)}_{k,\psi}(\omega,\vec{p}) = |\vec p|  \rho^{(Z)}_{k,\psi}(p^2) \, .
\end{equation}
We also note that the spectral function $\rho^{(C)}_{k,\psi}(\omega,\vec{p})$ is an even function of $\omega$ while  $\rho^{(A)}_{k,\psi}(\omega,\vec{p})$ and $\rho^{(B)}_{k,\psi}(\omega,\vec{p})$ are odd,
\begin{align}
\rho^{(C)}_{k,\psi}(-\omega,\vec{p}) &=  \rho^{(C)}_{k,\psi}(\omega,\vec{p})\, , \\
\rho^{(A)}_{k,\psi}(-\omega,\vec{p}) &=  -\rho^{(A)}_{k,\psi}(\omega,\vec{p})\, , \nonumber\\
\rho^{(B)}_{k,\psi}(-\omega,\vec{p}) &=  -\rho^{(B)}_{k,\psi}(\omega,\vec{p})\, . \nonumber
\end{align}
The Lehmann representation of the retarded propagator is given by
\begin{align}\label{lehmann}
G_{k,\psi}(\omega,\vec{p})=-\int_{-\infty}^{\infty}
d\omega'\frac{\rho_{k,\psi}(\omega',\vec{p})}{\omega'-\omega-i\epsilon}\, .
\end{align}
It can be used to derive various sum rules for the spectral functions as discussed in Sec.~\ref{sec:sumrules} below.

In the following we will set the spatial external momentum to zero for simplicity in this first study, $\vec{p}=0$, which implies $ G^{(A)}_{k,\psi}(\omega,0) \equiv 0$ and $ \rho^{(A)}_{k,\psi}(\omega,0) \equiv 0$ as well, and drop the corresponding second argument in all momentum dependent functions. The quark propagator can then be decomposed as
\begin{align}
G_{k,\psi}(\omega)=G_{k}^+(\omega)\Lambda_+ +G_{k}^-(\omega)\Lambda_-\,,
\end{align}
with $\Lambda_\pm=(1\pm\gamma_0)/2$ and
\begin{align}
G_{k}^\pm(\omega)=\frac{1}{2}\tr \big(G_{k,\psi}(\omega)\Lambda_\pm\big)\, .
\end{align}
With these we define the associated quark and anti-quark spectral functions,
\begin{align}
\rho_{k}^\pm(\omega) =\mp \frac{1}{\pi}\Im G_{k}^\pm(\omega) \, ,
\end{align}
such that $\rho_{k}^+(-\omega)=\rho_{k}^-(\omega)$. These are then related to the previously defined quark spectral functions by
\begin{align}
\label{eq:relation_spectral}
\rho_{k}^+(\omega)+\rho_{k}^-(\omega)=2\rho^{(C)}_{k,\psi}(\omega),\\
\rho_{k}^+(\omega)-\rho_{k}^-(\omega)=2\rho^{(B)}_{k,\psi}(\omega), \nonumber
\end{align}
and can be expressed in terms of the dressing functions of the two-point function as
\begin{align}
\rho_{k}^+(\omega)&=\frac{1}{\pi}
\frac{\Im C_k-\Im B_k}{(\Re C_k-\Re B_k)^2+(\Im C_k-\Im B_k)^2}\, , \\
\rho_{k}^-(\omega)&=\frac{1}{\pi}
\frac{\Im C_k+\Im B_k}{(\Re C_k+\Re B_k)^2+(\Im C_k+\Im B_k)^2}\, .
\end{align}

\section{Numerical implementation}\label{sec:numeric_implementation}

The flow equations for the effective potential, Eq.~(\ref{eq:flow_pot}), and for the dressing functions of the quark two-point functions, Eq.~(\ref{eq:flow_eq_2PF}), are solved using two different methods: the grid method and the Taylor method. Both methods use the same input parameters which are summarized in Tab.~\ref{tab:parameters}. These parameters are chosen such as to reproduce physical vacuum values for the pseudo-particle masses and the pion decay constant in the IR, see Tab.~\ref{tab:IR_values}. We note that there are small differences between the IR values obtained from the grid method and the Taylor method which, however, will not play any role in the following since we will only focus on qualitative differences in the results from these two methods, see also \cite{Pawlowski2014} for a comparison of different numerical implementations.

\begin{table}[h]
	\centering
	\begin{tabular}{C{1.3cm}|C{1.3cm}|C{1.1cm}|C{1.3cm}|C{1.3cm}}
		$\Lambda$/MeV & $m_\Lambda/\Lambda$ & $\lambda_\Lambda$ & $c/\Lambda^3$ &  $h$ \\
		\hline
		1000 & 0.794 & 2.00 & 0.00175 & 3.2 \\
	\end{tabular}
	\caption{Parameter set chosen for the quark-meson model, cf.~Eq.~(\ref{eq:gamma}).}
	\label{tab:parameters}
\end{table}

\begin{table}[h]
	\centering
	\begin{tabular}{C{1.6cm}|C{1.5cm}|C{1.5cm}|C{1.5cm}}
		$\sigma_0\equiv f_\pi$ & $m_\pi$ & $m_\sigma$ &  $m_\psi$  \\
		\hline
		93.5 MeV & 138 MeV & 509 MeV & 299 MeV \\
		90.1 MeV & 139 MeV & 534 MeV & 288 MeV
	\end{tabular}
	\caption{Observables obtained in the vacuum at an IR scale of $k_{\text{IR}}=40$~MeV when using the grid method (first row) and the Taylor method (second row).}
	\label{tab:IR_values}
\end{table}

\subsection{Grid method}
The idea of the grid method is to discretize the field variable $\phi^2=\sigma^2+\vec\pi^2$ on a grid in field space, see\cite{Schaefer:2004en} for details. The flow equation for the effective potential then turns into a set of coupled ordinary differential equations which can be solved using standard methods. The global minimum $\sigma_0$ of the effective potential in the IR determines the expectation value of $\sigma$ which is identified with the pion decay constant, $\sigma_0\equiv f_\pi$, at this leading-order in the derivative expansion.

The flow equation for the dressing functions of the quark two-point function can then be solved using the scale-dependent effective potential as an input. Since the flow equation for the two-point function does not couple different grid points, it is sufficient to use only one grid point $\phi^2_i$ which corresponds to the location of the global minimum at some chosen scale $k$. In the following we are mostly interested in the IR and therefore choose $\phi^2_i=\sigma_{0,\text{IR}}^2$. We also note that the flow equation for the two-point function is solved down to $k=0$ by using an extrapolation of the flow of the effective potential for $k<k_{\text{IR}}=40$~MeV. The same technique is used for the Taylor method.

One of the advantages of the grid method is that it does not restrict the shape of the effective potential and therefore allows for an almost arbitrary order of mesonic self-interactions (limited only by the number of grid points). In particular, possible secondary minima of the potential can be resolved which allows to study first-order phase transitions straightforwardly.

One possible issue with the grid method arises when solving the flow equation for the retarded two-point function. As discussed in App.~\ref{app:defs}, the $k$-integration of the flow equation for the imaginary part of the two-point function, which is closely connected to the spectral function, collapses to a sum over a few scales $k_i$ due to the appearance of Dirac delta functions. This means that for a given energy $\omega$, the spectral function may receive only a single contribution from some intermediate scale $k>k_{\text{IR}}$. Since the flow equation is always evaluated at the IR minimum $\phi^2_i=\sigma_{0,\text{IR}}^2$, also at $k>k_{\text{IR}}$, this contribution does not correspond to the actual scale-dependent minimum of $U_k(\phi^2)$. We will show, however, that the difference in the IR between spectral functions obtained from either grid or Taylor method (where this problem does not occur) is reasonably small.

\subsection{Taylor method}
There are different versions of the Taylor method available in the literature. The general idea is to use an expansion in the form of a Taylor series around some value of the field $\rho\equiv\phi^2$. The chosen expansion point may be constant, see for example \cite{Helmboldt2014}, or it may be scale dependent as discussed in the following. We first write the effective potential as
\begin{align}
U_k(\rho,\sigma)=\sum_{n=0}^{K}\frac{1}{n!}a_{n,k}(\rho-\rho_{0,k})^n - c\sigma,
\end{align}
where the expansion point $\rho_{0,k}$ is the scale-dependent minimum and we choose $K=5$. We use the same ansatz for the flow equation of the effective potential,
\begin{align}
\partial_kU_k(\rho)=\sum_{n=0}^{K}\frac{1}{n!}\partial_kU^{(n)}_k(\rho_{0,k})(\rho-\rho_{0,k})^n,
\end{align}
where $\partial_kU^{(n)}_k(\rho_{0,k})$ denotes the $n$-th derivative of the flow equation for the effective potential with respect to $\rho$, evaluated at $\rho_{0,k}$. From these two equations one can obtain flow equations for the coefficients $a_n$ and for $\rho_{0,k}$, which are given by
\begin{align}
\partial_k a_{n,k}&=\partial_kU^{(n)}_k(\rho_{0,k})+a_{n+1,k}\partial_k\rho_{0,k}.
\end{align}
When using that $\left.\partial_\sigma U_k(\rho,\sigma)\right|_{\sigma=\sqrt{\rho_{0,k}}}=0$, we find $a_{1,k}=c/(2\sqrt{\rho_{0,k}})$. By taking the RG-scale derivative of this relation one can express the flow equation for the scale-dependent minimum in terms of the flow equation for $a_{1,k}$, which gives
\begin{align}
\partial_k\rho_{0,k}&=-\frac{\partial_kU^{(1)}_k(\rho_{0,k})}{a_{2,k}+c/(4\rho_{0,k}^{3/2})}.
\end{align}

The same Taylor expansion can also be used for the mesonic two-point functions, see e.g.~\cite{Kamikado2014}. For the quark two-point function, which in the UV contains a term $h\sigma$, we will use a Taylor expansion in terms of $\sigma_{0,k}$ instead of $\rho_{0,k}$. The quark two-point function can then be written as
\begin{align}
\Gamma^{(2)}_{\psi,k}(\sigma)=\sum_{n=0}^{L}\frac{1}{n!}c_{n,k}(\sigma-\sigma_{0,k})^n,
\end{align}
where we use $L<5$. We can make a similar ansatz for the flow equation,
\begin{align}
\partial_k\Gamma^{(2)}_{\psi,k}(\sigma)=\sum_{n=0}^{L}\frac{1}{n!}\partial_k\Gamma^{(2),(n)}_k(\sigma_{0,k})(\sigma-\sigma_{0,k})^n,
\end{align}
where $\partial_k\Gamma^{(2),(n)}_k(\sigma_{0,k})$ denotes the $n$-th derivative of the flow equation for the two-point function with respect to $\sigma$, evaluated at $\sigma_{0,k}$. From these two equations we can obtain flow equations for the coefficients $c_{n,k}$ which are given by
\begin{align}
\label{eq:flow_Taylor}
\partial_k c_{n,k}&=\partial_k\Gamma^{(2),(n)}_k(\sigma_{0,k})+c_{n+1,k}\partial_k\sigma_{0,k}.
\end{align}

This Taylor method has the advantage that it always uses an expansion around the scale-dependent minimum of the effective potential. In this way, the two-point function only receives contributions that correspond to the local minimum at a given scale, in contrast to the grid method where the contributions are in general not evaluated at the global minimum for intermediate RG scales $k$. In the next sections we will present results obtained from both methods and discuss their differences.

\section{Masses and dressing functions $B_k(\omega)$ and $C_k(\omega)$}\label{sec:results1}
\begin{figure}[b]
	\includegraphics[width=0.49\textwidth]{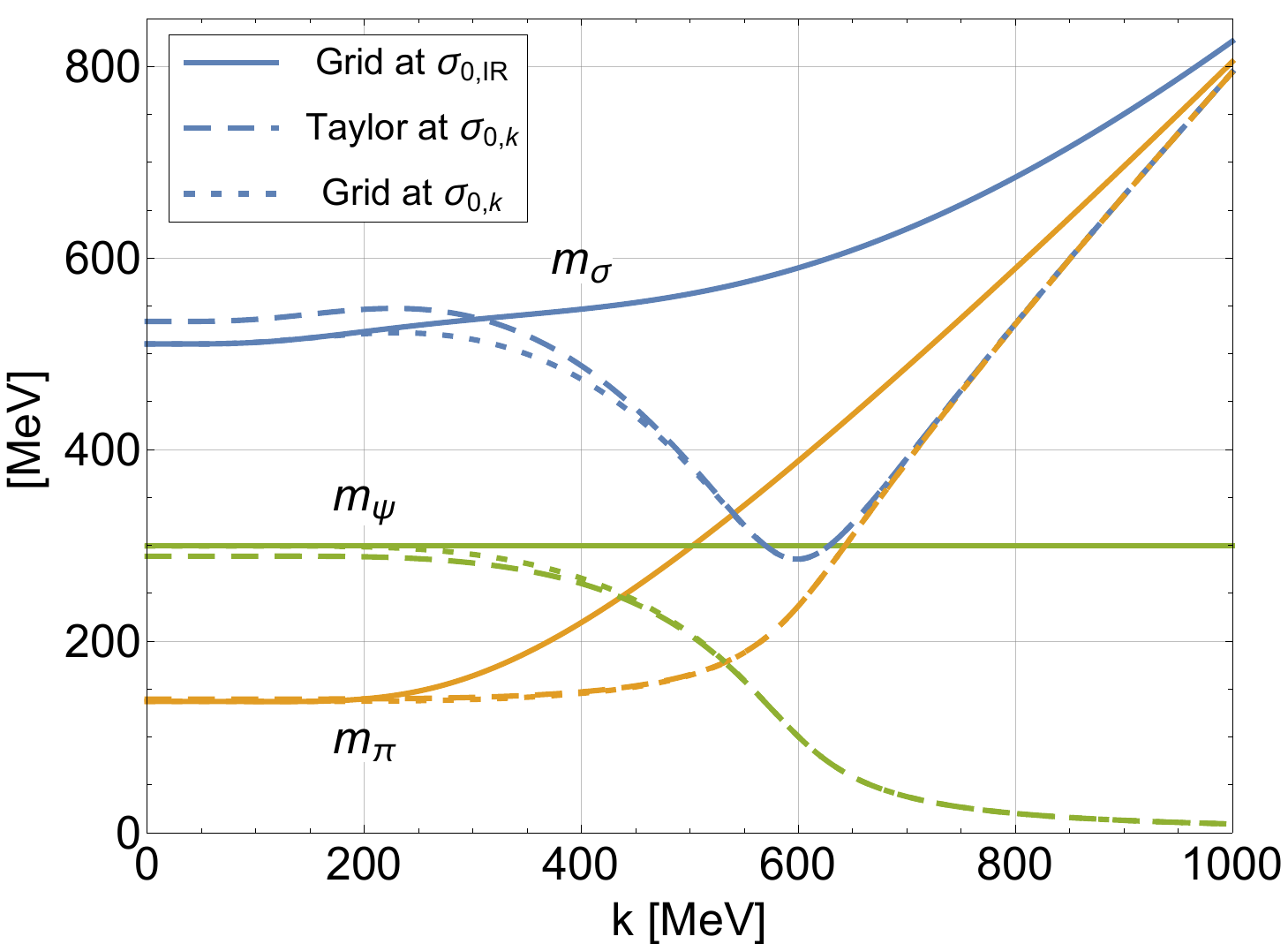}
	\caption{Flow of the Euclidean (curvature) masses $m_{\sigma,k}$ (blue), $m_{\pi,k}$ (yellow) and $m_{\psi,k}$ (green), cf.~Eq.~(\ref{eq:masses}), as obtained from the effective potential $U_k(\rho)$ using the grid method with $\rho=\rho_{0,\text{IR}}$ (solid lines), the Taylor method with $\rho=\rho_{0,k}$ (dashed lines) and the grid method with $\rho=\rho_{0,k}$ (dotted).}
	\label{fig:masses_pot}
\end{figure}

We will first study the flow of the Euclidean (curvature) masses as obtained from the effective potential using the grid method and the Taylor method, which are then used as input for the flow equation of the two-point function, see Fig.~\ref{fig:masses_pot}. Explicitly we have
\begin{align}
\label{eq:masses}
m_{\pi,k}^2=2U_k',\quad m_{\sigma,k}^2=2U_k'+4 U_k''\phi^2,\quad m_{\psi,k}^2=h^2\phi^2\,.
\end{align}
When evaluated at the scale-dependent global minimum of the effective potential, $\sigma_{0,k}$, these expressions represent the Euclidean curvature masses. When using the grid method as implemented in this work, however, the flow equation for the two-point function is always evaluated at the IR minimum $\sigma_{0,IR}$ while the Taylor method uses the scale-dependent minimum $\sigma_{0,k}$. The scale-dependent masses can of course also be obtained when using the grid method and evaluating Eq.~(\ref{eq:masses}) at $\sigma_{0,k}$. The masses then essentially agree with the ones obtained from the Taylor method, see Fig.~\ref{fig:masses_pot}.

We note that the quark mass obtained from the grid method using a fixed value of $\phi^2=\rho_{0,\text{IR}}$ is constant while when using the Taylor method the quark mass is almost zero in the UV and then significantly increases at the chiral symmetry breaking scale of $k\approx 600$~MeV where also the pion and the sigma mass become different. The masses obtained from both methods agree reasonably well in the IR, i.e.~up to the small deviations recorded in Tab.~\ref{tab:IR_values} which could be compensated by a small readjustment of the UV parameters.

\begin{figure}[b]
	\includegraphics[width=0.49\textwidth]{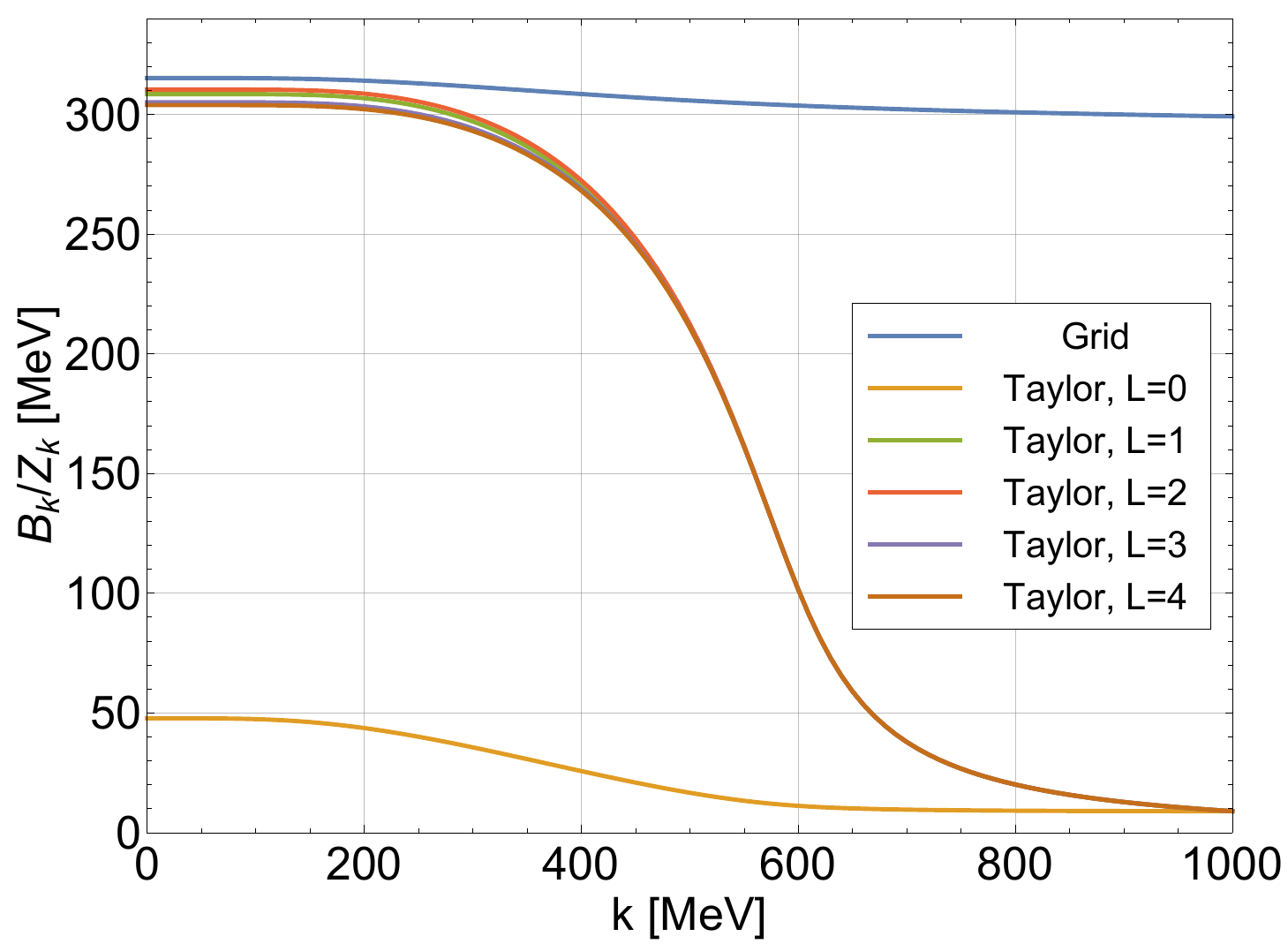}
	\caption{The mass dressing function $B_k$ divided by the renormalization function $Z_k$ (see Eq.~(\ref{eq:Z_k})), both at $\omega=0$ (and $\vec p=0$) where they are real, vs.~the scale $k$ obtained from the grid method and the Taylor method for different orders $L$.}
	\label{fig:flow_B}
\end{figure}

We will now study the flow of the mass dressing function $B_k$ divided by the renormalization function $Z_k$, see Eq.~(\ref{eq:Z_k}), of the retarded quark two-point function, as introduced in Eq.~(\ref{eq:Gamma2_R}). At $\omega = 0$, where they are both real, their ratio is shown in Fig.~\ref{fig:flow_B} where we compare the result from the grid method with those from the Taylor method for different expansion orders $L$.  When using the grid method, we have $B_{k=\Lambda}=h\sigma_{0,\text{IR}}=299$~MeV while for the Taylor method we have $B_{k=\Lambda}=h\sigma_{0,\text{UV}}=8.9$~MeV. Despite this large difference in the UV, both methods give approximately the same result in the IR, except for the Taylor method with $L=0$. Note that the Taylor method at order $L=0$ here produces a result that would agree with that from the grid method, if the scale-dependent minimum $\sigma_{0,k}$ was used in the latter in the place of the IR minimum $\sigma_{0,\text{IR}}$ that appears for example in the derivatives of the effective potential and in the initial condition for $\Gamma^{(2)}$ which contains $h\sigma$, see Eq.~(\ref{initialB}). Such a simple modification of the grid method would neglect contributions to the flow that arise from the $k$ dependence of the gliding minimum $\sigma=\sigma_{0,k}$ on the other hand. By comparison with the Taylor results, where these contributions are contained in the second term in Eq.~(\ref{eq:flow_Taylor}) and are thus absent at $L=0$, we see that they are by no means negligible but contribute substantially to the dynamical mass generation. Already the $L=1$ result is close to those from the higher orders in the Taylor expansion, however, and hence captures the main effect. We also note that the flow of the ratio $B_k/Z_k$ in the static limit resembles the flow obtained for the quark mass parameter in Fig.~\ref{fig:masses_pot} for the higher Taylor orders quite well.

\begin{figure}[t]
	\includegraphics[width=0.49\textwidth]{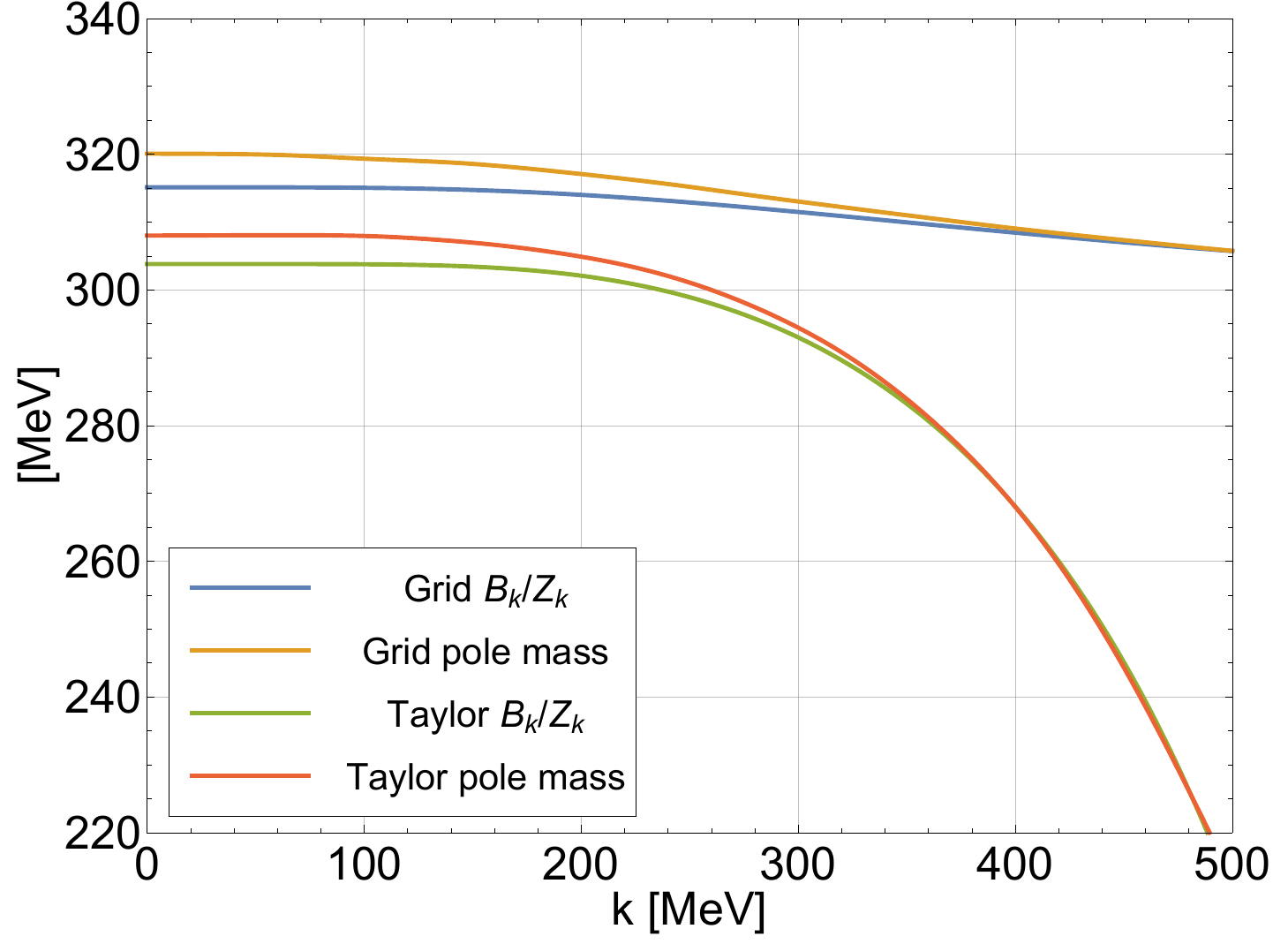}
	\caption{The flow of the mass dressing function $B_k$ divided by the renormalization function $Z_k$ at $\omega=0$ and of the quark pole mass as obtained by using the grid method and the Taylor method with $L=4$.}
	\label{fig:flow_pole_mass_Bk2}
\end{figure}

\begin{figure}[t]
	\includegraphics[width=0.49\textwidth]{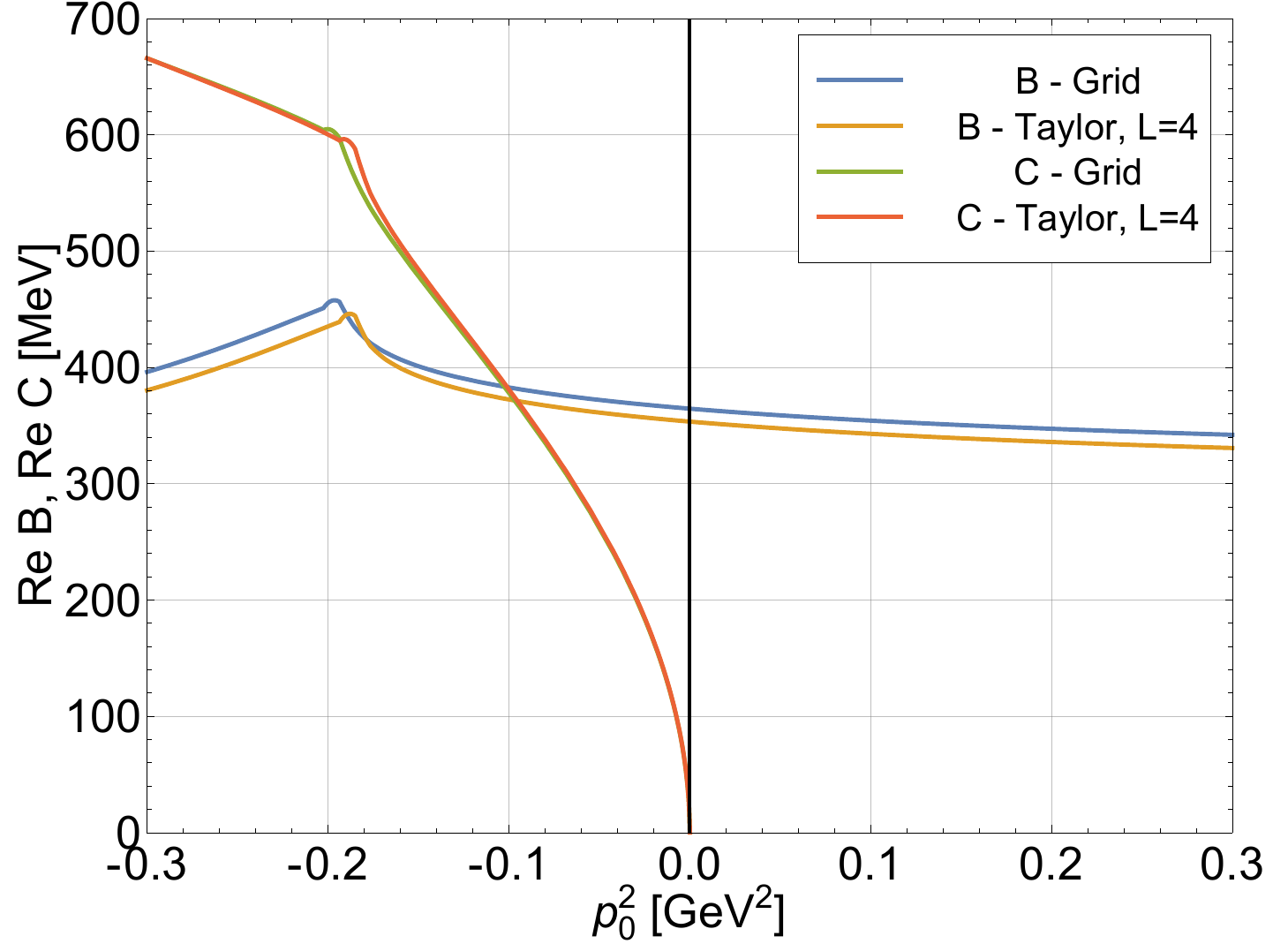}
	\caption{The real part of the functions $B_k$ and $C_k$ for $k\to 0$ as obtained by using the grid method and the Taylor method with $L=4$. We note that the quark mass obtained from the potential is the same as the value for $B$ in the limit $p_0\rightarrow \infty$.}
	\label{fig:ReB_M_E_1}
\end{figure}

In Fig.~\ref{fig:flow_pole_mass_Bk2} we compare the flow of $B_k/Z_k$ at $\omega = 0$  with the flow of the scale-dependent quark pole mass $m_{\psi,k}^P$ which is obtained as the (lowest) value of $\omega$ that solves $B_k(\omega)=C_k(\omega)$ in the range where both dressing functions are still real and thus in the IR describes the stable single-particle contribution to the quark propagator here. We find that the pole mass starts to deviate from $B_k/Z_k$ for $k\lesssim 400$~MeV and that the pole mass is several MeV larger than the $\omega=0$ value of $B_k/Z_k$ in the IR. In general, a difference between the pole mass and the mass function $B_k/Z_k$ at $\omega=0$ is to be expected since $B_k/Z_k$ has a non-trivial $\omega$ dependence which is seen in Fig.~\ref{fig:ReB_M_E_1} where we plot the real parts of $B$ and $C$ (in the IR) over $p_0^2$. With increasing negative values of $p_0^2 = p^2$ here, i.e.~deeper in the timelike region, both $B$ and $C$ increase as long as they remain real until nonzero imaginary parts develop at their peak position as discussed in the next section.

For positive $p_0^2 = p^2$, i.e.~in the spacelike region, $C$ becomes pure imaginary because it is defined as $C(z) = z Z(z)$ with $z= ip_0$ in the Euclidean domain, while $B$  monotonically decreases with the Euclidean $p^2$. Because the explicit flow of the two-point function vanishes when $p^2 \gg \Lambda^2 $, $B$ approaches $g\sigma_{0,\text{IR}} = m_\psi $ in Tab.~\ref{tab:IR_values}. In the grid method the IR value of $\sigma_0$ is the same as its UV value which means that also $B_k$ remains at its UV initial condition, Eq.~(\ref{initialB}), so that $B_k$ does not flow at all in this case. In the Taylor method one starts with a small value for $B_{k=\Lambda} = h\sigma_{0,\text{UV}}$, as mentioned above, but the flow of $\sigma_{0,k}$ leads to an implicit residual flow also for $B_k$ at asymptotically large momenta as seen explicitly in Eq.~(\ref{eq:flow_Taylor}) (for $L \ge 1$). This contribution ensures that $B_k$ eventually approaches the corresponding infrared value of the quark mass parameter in Tab.~\ref{tab:IR_values} with the Taylor method as well.

\begin{figure*}[t]
	\includegraphics[width=0.49\textwidth]{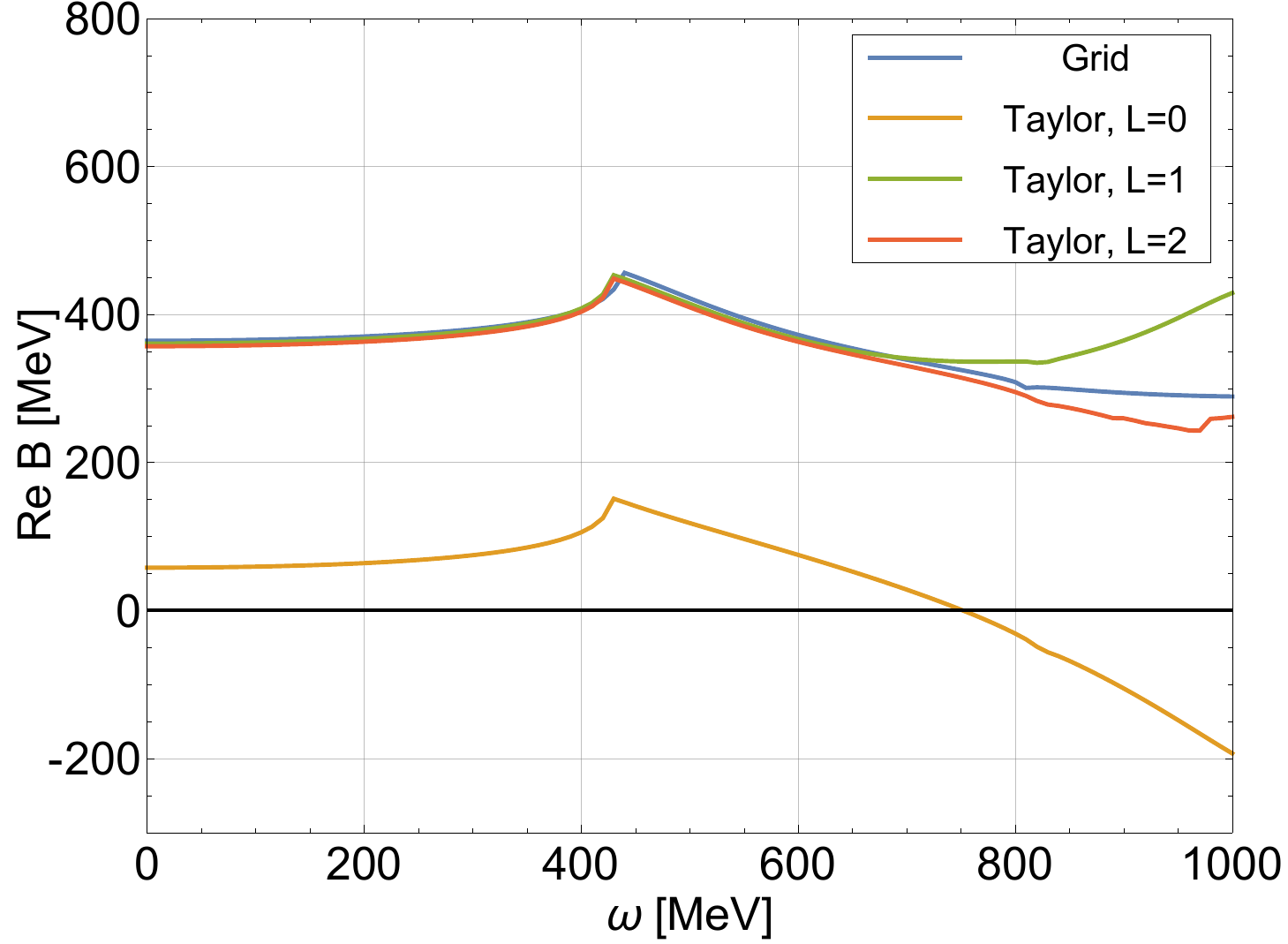}
	\includegraphics[width=0.49\textwidth]{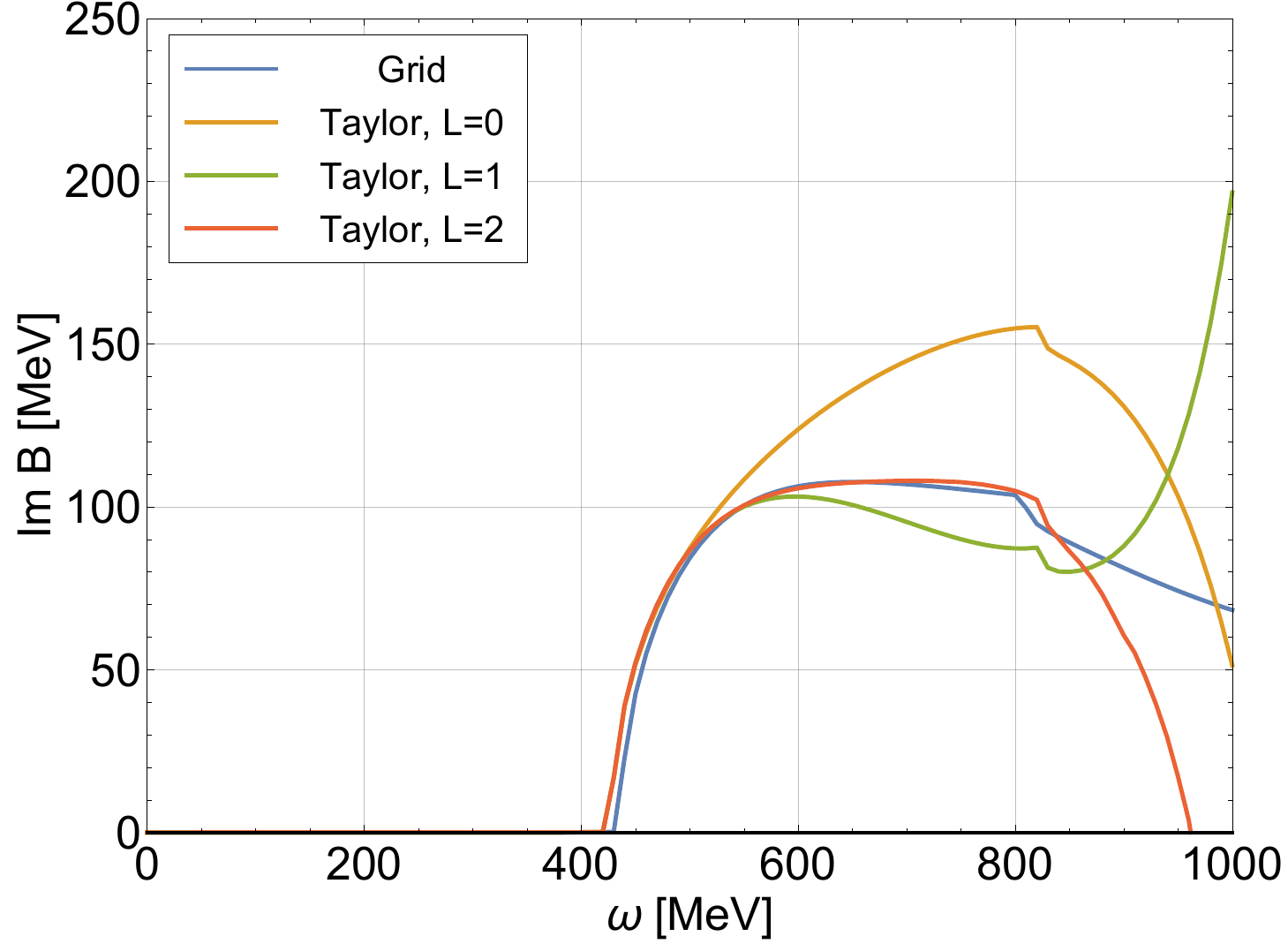}
	\caption{Real (left) and imaginary (right) parts of $B(\omega) =B_k(\omega)$, $k\to 0$ from grid or Taylor method at different orders $L$.}
	\label{fig:B_omega}
\end{figure*}

\begin{figure*}[t]
	\includegraphics[width=0.49\textwidth]{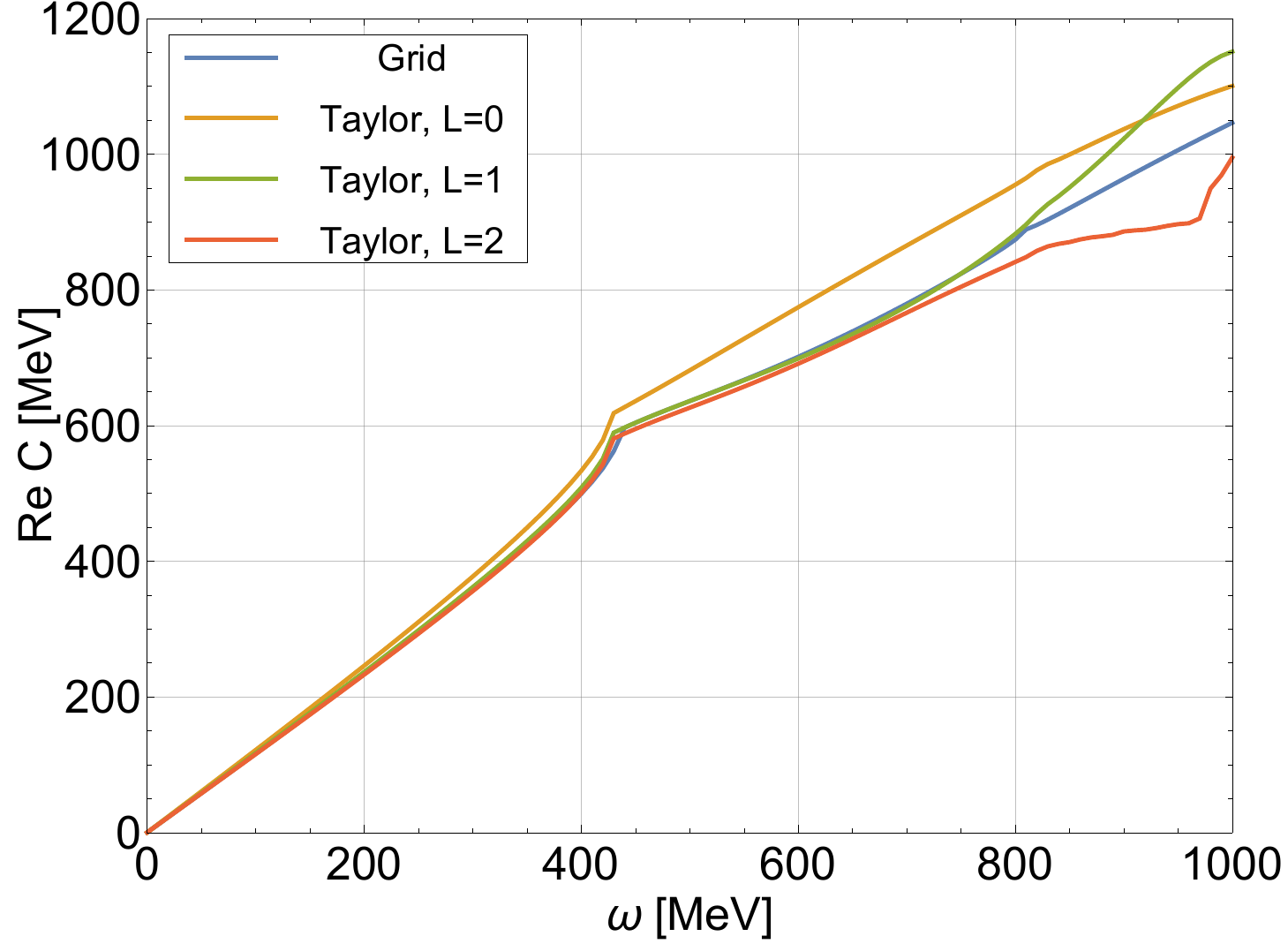}
	\includegraphics[width=0.49\textwidth]{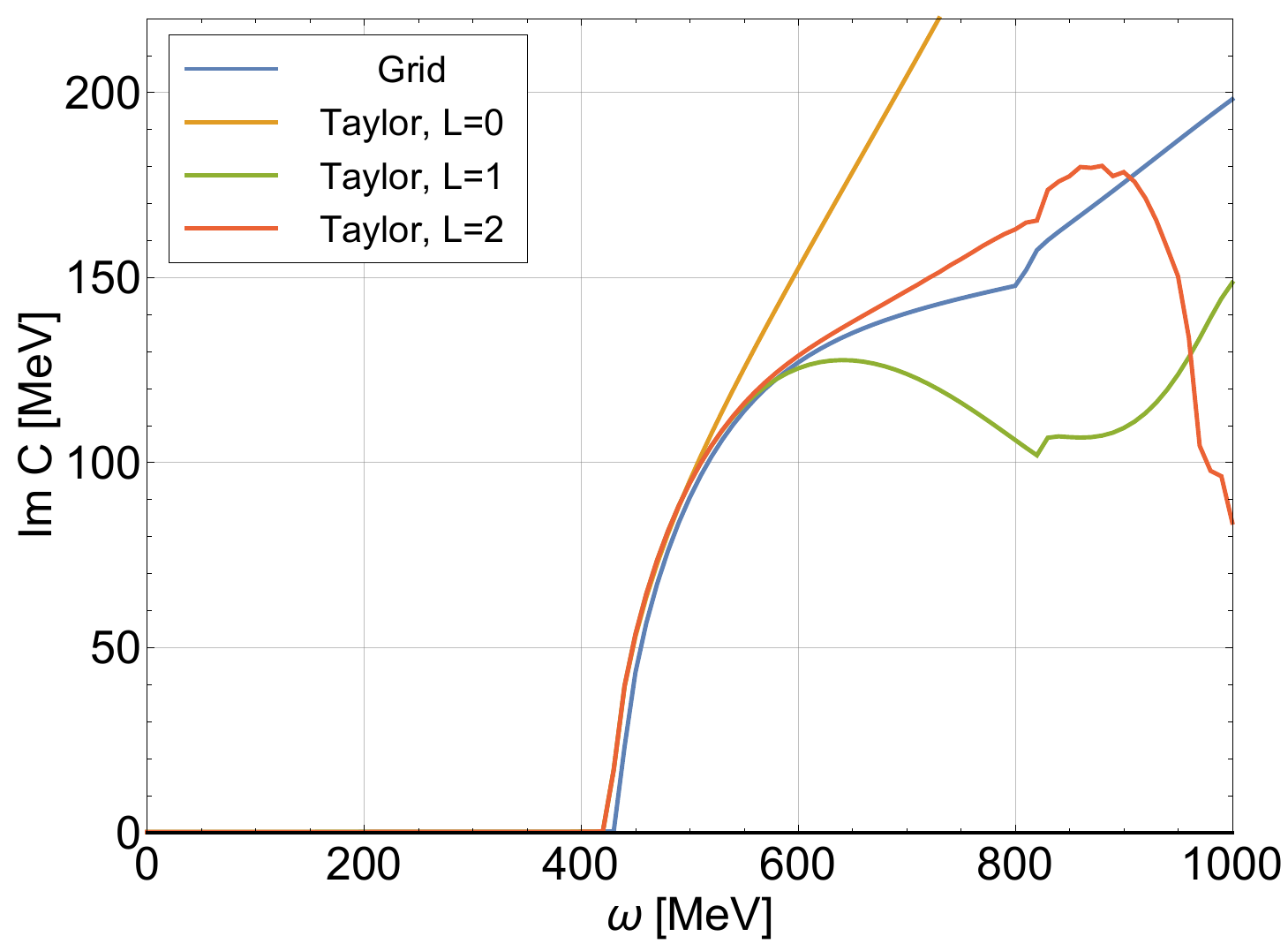}
	\caption{Real (left) and imaginary (right) parts of $C(\omega)= C_k(\omega)$, $k\to 0$ from grid or Taylor method at different orders $L$.}
	\label{fig:C_omega}
\end{figure*}

\section{Quark Spectral Function}\label{sec:results2}

We now turn to the flow of the different quark spectral functions: $\rho_{k,\psi}^{(B)}$, $\rho_{k,\psi}^{(C)}$, $\rho_{k,+}$ and $\rho_{k,-}$. They all depend on the dressing functions $B_k(\omega)$ and $C_k(\omega)$. The infrared results, as obtained with $k\to 0$ from either grid or Taylor method, for real and imaginary parts of $B(\omega)$ are shown in Fig.~\ref{fig:B_omega}, the corresponding ones for $C(\omega)$ in Fig.~\ref{fig:C_omega}.

The shapes of $B(\omega)$ and $C(\omega)$ which are also reflected in the spectral functions can be explained by considering the different particle processes that can occur within our framework. These processes can already be read off from the diagrammatic representation of the flow equation for the two-point function, see Fig.~\ref{fig:flow_Gamma2}, and are given by
\begin{align}
\psi^* &\rightarrow\psi+\pi && \text{for} \quad \omega\ge E_{\psi}+E_{\pi},\\
\psi^* &\rightarrow\psi+\sigma && \text{for} \quad \omega\ge E_{\psi}+E_{\sigma},\\
\bar\psi^* &\rightarrow\bar\psi+\pi && \text{for}\quad  \omega\ge E_{\psi}+E_{\pi},\\
\bar\psi^* &\rightarrow\bar\psi+\sigma && \text{for} \quad \omega\ge E_{\psi}+E_{\sigma},
\end{align}
where $\psi^*$ denotes an off-shell quark with energy $\omega$ and the other energies represent IR values. The process $\psi^* \rightarrow\psi+\pi$ for example describes a quark with energy $\omega$ that can `decay' into an on-shell quark with energy $E_{\psi}$ and a pion with energy $E_{\pi}$. These processes allow for a clear interpretation of the shape of the real and in particular of the imaginary part of $B(\omega)$ and $C(\omega)$.

The real part of $B(\omega)$ starts with a small positive slope at small energies and monotonously increases up to the first threshold at $\omega=E_{\psi}+E_{\pi}\approx 420$~MeV where the decay channel into a quark and a pion opens up. A second, but smaller change in the real part is visible at the second threshold at $\omega=E_{\psi}+E_{\sigma}\approx 820$~MeV where the process $\psi^*\rightarrow\psi+\sigma$ becomes possible. The imaginary part of $B(\omega)$ stays at zero below the first threshold energy $\omega=E_{\psi}+E_{\pi}$ since no decay channels are available and then starts to rise quickly at the quark-pion threshold. The quark-sigma process gives rise to a small kink in the imaginary part at $\omega = E_{\psi}+E_{\sigma}$.Up to these energies the result obtained from the grid method is in good agreement with the results from the Taylor method for $L\ge 1 $. For higher Taylor orders $L\ge 3$ we observe numerical difficulties at large energies. The behavior of $C(\omega)$ is is analogous: below the quark-pion threshold it stays real, with a rapidly increasing imaginary part starting there and further kinks in real and imaginary parts at the quark-sigma threshold. The leading behavior at small $\omega$ is given by  $ C(\omega) = Z \omega + \mathcal O(\omega^2)$ with $Z = Z_{k=0}(0) \approx 1.16$. Note that both $B(z)$ and $C(z)$ are analytic functions at $z = 0$ as can be seen from their Lehmann representation and the fact that their spectral functions have no support there as we will discuss next.

\begin{figure}[t]
	\includegraphics[width=0.49\textwidth]{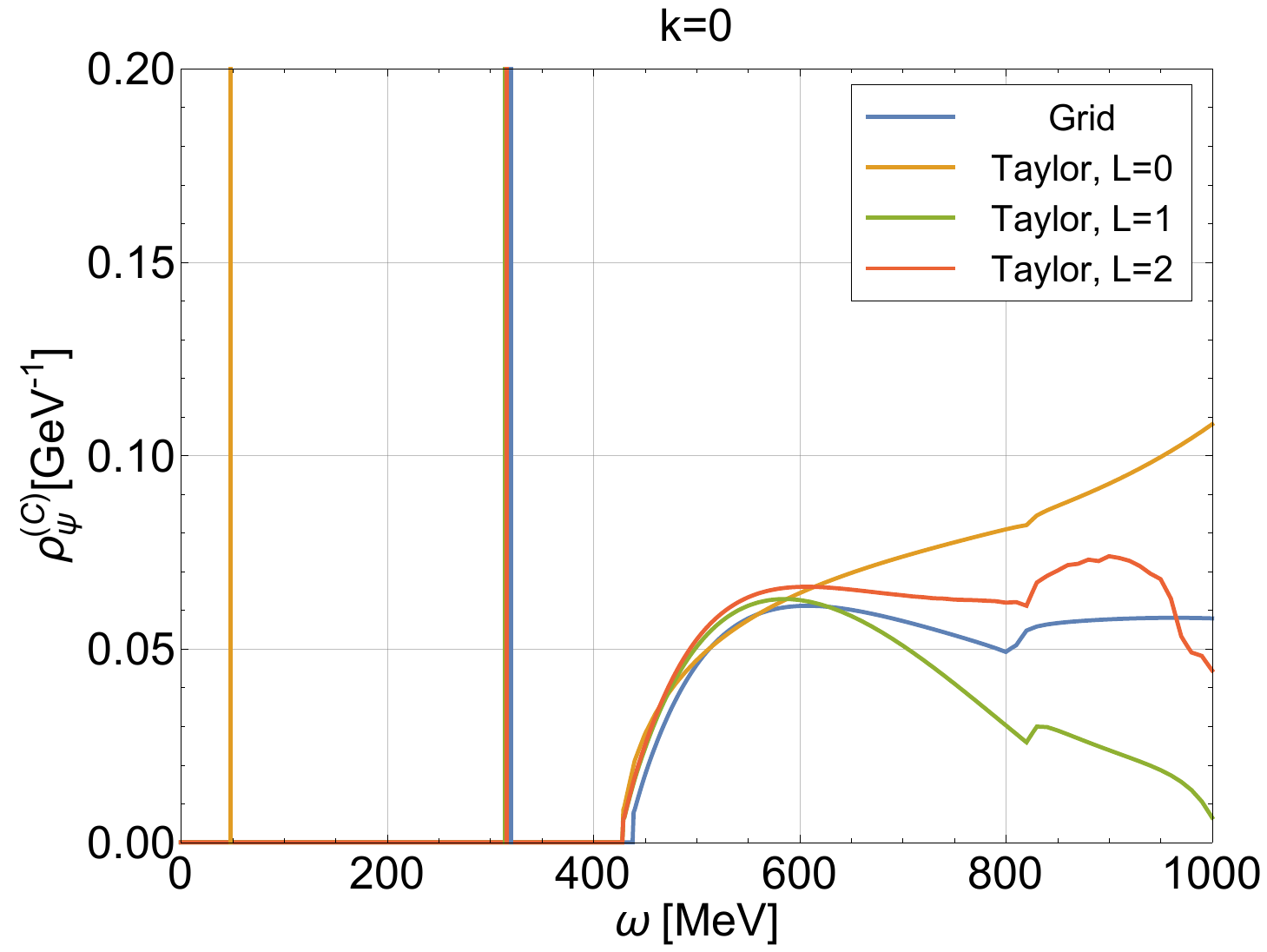}
	\caption{The quark spectral function $\rho^{(C)}_{\psi}(\omega)$ as a function of $\omega$ at $k=0$ as obtained by using the grid method as well as the Taylor method for different orders $L$.}
	\label{fig:rho_C}
\end{figure}

We now turn to the quark spectral function $\rho^{(C)}_{k,\psi}(\omega)$. It starts in the UV as a simple delta function at the UV quark mass parameter with strength $1/2$, cf.~Eq.~(\ref{initialSpec}), and flows with $k\to 0$ towards the infrared results shown in Fig.~\ref{fig:rho_C} as obtained from the grid method and the Taylor method. Although the UV values are very different, cf.~Eq.~(\ref{eq_UV_values_coeffs}), the spectral functions agree well in the IR, in particular for higher Taylor orders like $L=2$. While the delta peak that is connected to the quark pole mass with the Taylor method moves from $\omega\approx 9$~MeV in the UV up to a value of $\omega\approx 316$~MeV in the IR, the pole mass obtained from the grid method changes with the flow only from $\omega\approx 299$~MeV in the UV to $\omega\approx 320$~MeV in the IR. The remaining discrepancy of about $4$~MeV between the IR pole masses could in principle be compensated by a slight readjustment of the UV parameter as mentioned above. These single particle contributions at the physical mass are defined as the solutions to $B(\omega)= C(\omega)$ and indicated by the vertical lines in the figures.

The difference between curvature masses, obtained from the effective potential and the physical pole masses has also been observed for mesons, see e.g.~\cite{Tripolt2014}. While the two need not be the same of course, but differ whenever one has frequency dependent renormalization effects as those in the ratio $B(\omega)/Z(\omega)$ here, and explicitly demonstrated in Fig.~\ref{fig:ReB_M_E_1} above, the size of the difference is generally determined by the relative distance of the closest singularity above the single-particle pole. As such the effect observed especially for the pions in the previous LPA studies was too large. To reduce this artifact one has to go beyond the leading order in the derivative expansion by least including scale-dependent wavefunction renormalization factors \cite{Strodthoff:2016pxx}. We should therefore be prepared that analogous improvements can also lead to similar quantitative changes here. As a next step towards a fully self-consistent calculation one should therefore extract these scale-dependent wavefunction renormalization factors and feed them back into the calculation by iteration in future as well.  

We have already noted that the Taylor method at the leading order, with $L=0$, misses an important qualitative effect. Higher orders on the other hand appear to converge quickly towards the grid result. We then generally have quite compelling agreement between both methods in the IR (provided the external frequency $\omega $ stays well below the cutoff scale $\Lambda$). The remaining discrepancies can in fact be considered as an indication of systematic uncertainties. The particular advantage of the Taylor-expansion method is that it provides a direct and more intuitive understanding of the evolution of the spectral function with the flow at intermediate scales $k$. To achieve this with the grid method one would best stop the flow at some intermediate scale $\bar k$ and study the behavior of the spectral functions in the fixed background with $\sigma = \sigma_{0, \bar k} $ given by the minimum of the effective potential at this intermediate scale $\bar k$. In the following we focus on the physical results  obtained at the end of the flow in the IR. Having established the equivalence of the methods there, we restrict to the grid results for clarity from now on.

\begin{figure}[t]
	\includegraphics[width=0.49\textwidth]{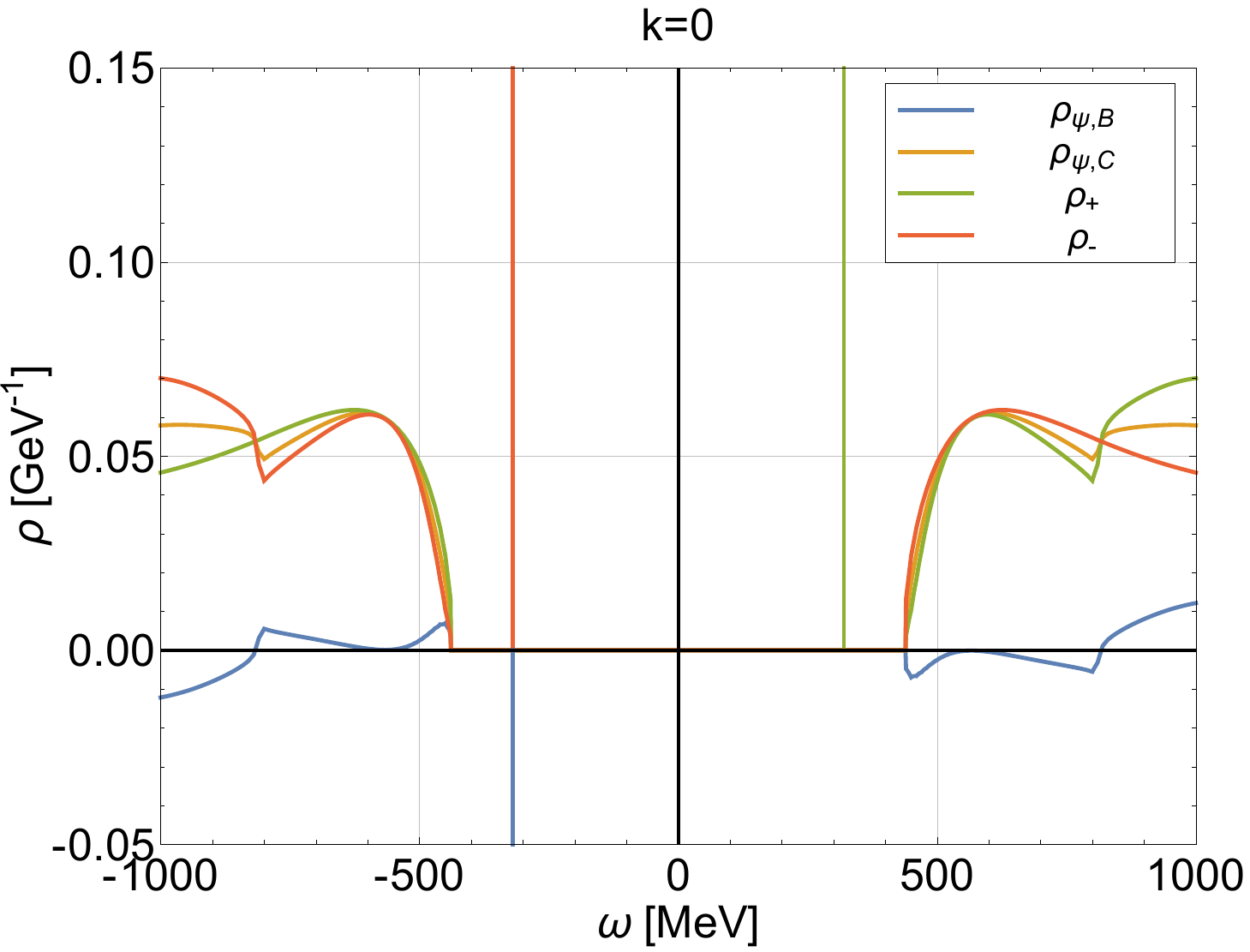}
	\caption{The spectral functions $\rho^{(B)}_{\psi}(\omega)$ and $\rho^{(C)}_{\psi}(\omega)$ together with the quark spectral function $\rho_{+}(\omega)$ and the anti-quark spectral function $\rho_{-}(\omega)$ in the IR as obtained with the grid method.}
	\label{fig:rho_C_etc}
\end{figure}

As a first check note that $\rho^{(C)}_{\psi}(\omega)$ is the positive distribution that it has to be in a theory with a positive state space, and one has the inequality
\begin{equation}
   \rho_\psi^{(C)}(\omega) \ge | \rho^{(B)}_{k,\psi}(\omega) | \, , 
\end{equation}
which is satisfied here as well. For the continuum contributions this can be seen explicitly in Fig.~\ref{fig:rho_C_etc} where we show  $\rho^{(B)}_{\psi}(\omega)$ and  $\rho^{(C)}_{\psi}(\omega)$ together with the quark and antiquark spectral functions, cf.~Eq.~(\ref{eq:relation_spectral}),
\begin{align*}
  \rho_{+}(\omega) &= \rho^{(C)}_{\psi}(\omega) + \rho^{(B)}_{\psi}(\omega) \;\;\mbox{ and} \nonumber \\
  \rho_{-}(\omega) &=  \rho^{(C)}_{\psi}(\omega) - \rho^{(B)}_{\psi}(\omega) \, ,
\end{align*}
which are therefore also both positive. The single-particle contributions in $\rho^{(B)}_{\psi}(\omega)$ and  $\rho^{(C)}_{\psi}(\omega)$ have the same magnitude. Consequently, the quark spectral function $\rho_{+}(\omega)$ only exhibits one delta peak at positive energies, representing a single quark, while the antiquark spectral function $\rho_{-}(\omega)$ only has a peak at negative energies for the single-antiquark states.

The continuum parts of $\rho^{(C)}_{\psi}(\omega)$ and  $\rho^{(B)}_{\psi}(\omega)$ related to the various one-to-two-particle processes add up in $\rho_{\pm}(\omega)$, which leads to an enhancement at large positive energies, while they are subtracted from one another  in $\rho_{-}(\omega)$ and hence suppressed there. The corresponding converse behavior follows with $\rho^{(B)}_{\psi}(-\omega) = - \rho^{(B)}_{\psi}(\omega)$ and $\rho^{(C)}_{\psi}(-\omega) = - \rho^{(C)}_{\psi}(\omega)$ for $\omega < 0$. The slightly negative contributions to $ \rho^{(B)}_{\psi}(\omega)$ in the range between the quark-pion and the quark-sigma threshold are likely to be an artifact of the present truncation.

\section{Sum Rules}\label{sec:sumrules}
The various sum rules as usual follow from expanding the Lehmann representation in Eq.~(\ref{lehmann}),
\begin{equation}
   G_\psi(\omega) = \int_{-\infty}^\infty \! d\omega' \,\frac{\rho_\psi(\omega')}{\omega-\omega'} \, ,
\end{equation}
for small and large $\omega$. For large $\omega $, this leads to
\begin{align}
  G_\psi(\omega) &=
  \int_{-\infty}^\infty \! d\omega' \,\frac{\rho_\psi(\omega')}{\omega}  \, \Big( 1 + \frac{\omega'}{\omega} + \cdots  \Big) \\
  & =  \gamma_0  \frac{1}{\omega} \int_{-\infty}^\infty \! d\omega' \rho^{(C)}_\psi(\omega') \label{SRaslargeo}\\
  &\hskip 1cm + \frac{1}{\omega^2} \int_{-\infty}^\infty \! d\omega' \, \omega' \rho^{(B)}_\psi(\omega')  \, + \, \mathcal O(1/\omega^3) \, , \nonumber
\end{align}
where we have used that the even moments of the odd function $\rho_\psi^{(B)}(\omega)$ and the odd moments of the even $\rho^{(C)}_\psi(\omega) $ both vanish. The leading order at large $\omega $ therefore corresponds to,
\begin{align}
  \int_{-\infty}^\infty \! d\omega \, \rho_\psi^{(B)}(\omega) &= 0 \, ,\\
  \int_{-\infty}^\infty \! d\omega \, \rho_\psi^{(C)}(\omega) &= \lim_{\omega\to\infty} \omega G_\psi^{(C)}(\omega)  = \lim_{\omega\to\infty} Z^{-1}(\omega^2) \, . \label{LOSRC}
\end{align}
In a renormalizable field theory, the integral of the spectral density usually diverges logarithmically and the right-hand side is then given by a formally ultraviolet divergent field renormalization constant. Here, we note that the flow of the two-point function generally vanishes for $\omega \to \infty$ and the leading-order sum rule for $\rho_{k,\psi}^{(C)}(\omega) $ consistent with the UV initial condition in Eq.~(\ref{initialC}) therefore reads
\begin{align}
\int_{-\infty}^\infty\! d\omega \,\rho^{(C)}_{k,\psi}(\omega) = 1. \label{LOSRC1}
\end{align}
It is satisfied exactly in the UV by definition, where the quark spectral function is given by
\begin{align}
\rho^{(C)}_{\Lambda,\psi}(\omega)=Y_\Lambda (\delta(\omega-m_{\psi,\Lambda})+\delta(\omega+m_{\psi,\Lambda}))
\end{align}
with $Y_\Lambda=1/2$. As fluctuations are included with integrating the flow down to some scale $k$, the weight of these free single-(anti)particle contributions gets reduced in favor of continuum contributions from processes that are possible due to the interactions with fluctuations included above this scale. Monitoring the sum rule (\ref{LOSRC1}) over the flow therefore provides a valuable test of the consistency of the procedure.

For this purpose, we evaluate all sum rules numerically up the the sum of the UV energies, $\Lambda_E=E_{\psi,\Lambda}+E_{\alpha,\Lambda}\approx 2324$~MeV, since at this scale the FRG flow gives the first contribution to the continuum and the strength of the Dirac delta peak starts to decrease. During the flow, the quark and meson energies decrease and the threshold of the continuum moves to smaller energies until it reaches its IR value. When applied to the quark spectral function obtained from the grid method for example this then yields in the IR
\begin{align}
\int_{-\Lambda_E}^{\Lambda_E} d\omega \:\rho^{(C)}_{k=0,\psi}(\omega) \approx 1.094\, .
\end{align}
Herein, we have included the delta functions from the single-particle contributions to the spectral function explicitly, with scale dependent pole mass $m_{\psi,k}^P$ and residue $Y_k$, by using
\begin{align}
\rho^{(C), \mathrm{pole}}_{k,\psi}(\omega)=Y_k (\delta(\omega-m_{\psi,k}^P)+\delta(\omega+m_{\psi,k}^P))
\end{align}
with
\begin{align}
Y_k=\frac{1}{2}\,  \big| (\partial_\omega C_k(\omega)-\partial_\omega B_k(\omega))|_{m_{\psi,k}}\big|^{-1} .
\end{align}
We find $2Y_{k=0}\approx 0.875$ which means that approximately $88\%$ of the sum rule are still provided by the single-particle and antiparticle contributions, which only leaves less than half of the sum rule for the continuum from the interactions here, at $T=\mu=0$. The fact that the sum rule remains satisfied numerically to a high degree all the way down to the infrared demonstrates the consistency of the FRG approach in that it also keeps the normalization intact, in addition to preserving Dirac and symmetry structures.

At next-to-leading order, the expansion in Eq.~(\ref{SRaslargeo}) corresponds to the energy-weighted sum rules for the quark propagator,
\begin{align}
  \int_{-\infty}^\infty \! d\omega \, \omega\, \rho_\psi^{(C)}(\omega) &= 0 \, ,\\
  \int_{-\infty}^\infty \! d\omega \,  \omega\, \rho_\psi^{(B)}(\omega) &= \lim_{\omega\to\infty} \omega^2 \,\frac{1}{4} \tr\; G_\psi(\omega) \label{NLOSRB} \\
  &= \lim_{\omega\to\infty} \omega^2 \, G^{(B)}(\omega) = \lim_{\omega\to\infty}
  \frac{B(\omega)}{Z^2(\omega^2)}\, . \nonumber
\end{align}
By the same argument as above, the flow for the two-point function vanishes in this the limit which is therefore here given by the UV initial conditions in Eqs.~(\ref{initialB}) and (\ref{initialC})  again. The energy-weighted sum rule for the quark propagator therefore here becomes
\begin{align}
  \int_{-\infty}^\infty\! d\omega \, \omega\, \rho^{(B)}_{k,\psi}(\omega)
  = m_{\psi,\Lambda}\, .
\end{align}
Our parameters for the grid method in Table \ref{tab:IR_values} imply for the quark mass at the UV cutoff scale $m_{\psi,\Lambda}=299$~MeV (the quark mass parameter $m_\psi $  is in fact scale-independent when using the grid code). For comparison, the numerical value of the energy-weighted integral over $\rho^{(B)}(\omega)$ in the infrared would correspond to  $m_{\psi,\Lambda}=324.5$~MeV. This sum rule is therefore slightly violated, but only within an error of about 9\%.

The leading-order sum rule (\ref{LOSRC}) for $\rho_\psi^{(C)}(\omega) $ and the next-to-leading-order sum rule (\ref{NLOSRB}) for $\rho_\psi^{(B)}(\omega) $ are determined by the purely perturbative behavior of the quark propagator. Higher moments of the spectral functions diverge (more than logarithmically) in the ultraviolet. The corresponding contributions to the propagator are suppressed by powers of $1/\omega^2$ relative to the leading ones at large $\omega $ and need to be obtained from the operator product expansion.

Negative moments on the other hand will converge more and more rapidly in the ultraviolet. As long as there are no contributions to the spectral function from massless excitations there will be no infrared divergences either. To obtain these moments one expands Eq.~(\ref{lehmann}) for small $\omega$,
\begin{align}
  G_\psi(\omega) &=
  - \int_{-\infty}^\infty \! d\omega' \,\frac{\rho_\psi(\omega')}{\omega'}  \, \Big( 1 + \frac{\omega}{\omega'} + \cdots  \Big) \\
  & =  - \int_{-\infty}^\infty \! d\omega' \,
  \frac{\rho^{(B)}_\psi(\omega')}{\omega'}
  \label{SRassmallo}\\
  &\hskip 1cm - \gamma_0 \, \omega \int_{-\infty}^\infty \! d\omega' \, \frac{\rho^{(C)}_\psi(\omega')}{{\omega'}^2}  \, + \, \mathcal O(\omega^2) \, , \nonumber
\end{align}
where we have again used that the even(odd) moments of $\rho_{\psi}^{(B)}(\omega)$($\rho_{\psi}^{(C)}(\omega)$) vanish. The non-vanishing ones then lead to the leading-order,
\begin{equation}
  \int_{-\infty}^\infty \! d\omega \,
  \frac{\rho^{(B)}_\psi(\omega)}{\omega} = -G_\psi(0) = \frac{1}{B(0)}\, ,
  \label{loSRso}
\end{equation}
and the next-to-leading-order sum rule,
\begin{align}
  \label{nloSRso}
  \int_{-\infty}^\infty \! d\omega \,
  \frac{\rho^{(C)}_\psi(\omega)}{\omega^2} &= -\lim_{\omega\to 0} \frac{1}{4}\, \frac{\tr\, \big( \gamma_0 G_\psi(\omega)\big)}{\omega}  \\
  &= - \lim_{\omega\to 0} \frac{G_\psi^{(C)}(\omega)}{\omega}  = \frac{Z(0)}{B^2(0)}\, .\nonumber
\end{align}
Both sides of these sum rules are now scale dependent.

For the leading-order sum rule we start in the UV with $1/B_{k=\Lambda}(0)=1/m_{\psi,k=\Lambda}\approx 3.344\cdot10^{-3}$\;$\text{MeV}^{-1}$ and the sum rule is trivially satisfied. More importantly, however, in the IR we find $1/B_{k=0}(0)\approx 2.744\cdot10^{-3}$\;$\text{MeV}^{-1}$, and this then compares very well with the numerical value of the integral in Eq.~(\ref{loSRso}) which gives $2.752\cdot10^{-3}$\;$\text{MeV}^{-1}$.

Also the next-to-leading-order sum rule is trivially satisfied in the UV, with  $Z_{k=\Lambda}(0)=1$ and $1/B_{k=\Lambda}^2(0)=1/m_{\psi,k=\Lambda}^2\approx 1.119\cdot10^{-5}$\;$\text{MeV}^{-2}$. In the IR, on the other hand, we have $Z_{k=0}(0)\approx 1.156$ and $Z_{k=0}(0)/B_{k=0}^2(0)\approx 8.706\cdot10^{-6}$\;$\text{MeV}^{-2}$, very close to the numerical value of the integral in Eq.~(\ref{nloSRso}) which gives $8.739\cdot10^{-6}~\text{MeV}^{-2}$ for comparison.

\section{Summary and Outlook}\label{sec:summary}

In this work we have presented a framework to calculate fermionic spectral functions within the Functional Renormalization Group approach. Our method is based on a recently developed technique for the calculation of bosonic spectral functions that uses a well-defined analytic continuation procedure from imaginary to real energies. The fact that no numerical analytic continuation method is needed represents a distinct advantage over other approaches that have to rely on numerical reconstruction techniques like the Maximum Entropy Method.

In the present study we applied this method to the quark-meson model and calculated quark spectral functions in the vacuum. The resulting flow equations for the real-time two-point functions have been solved numerically using two different methods: the grid method and the Taylor method. Both methods produce consistent results when the corresponding flow equations are integrated all way down to the infrared, with residual discrepancies serving as indications of the systematic uncertainties. Thereby, the grid method by and large produces the more stable results, while the Taylor method provides the more direct and intuitive interpretation of the full scale-dependence of the spectral functions during the flow.

In particular, we studied the flow of the quark mass as well as the quark and anti-quark spectral functions. The different particle processes which define the shape of the spectral functions as well as the consistency with various sum rules, derived from the Lehmann representation of the quark propagator, have been discussed.

Although we have limited ourselves to the vacuum and to vanishing spatial momenta in this first work on fermionic spectral functions with the FRG, our approach can be extended to finite temperature, finite chemical potential and finite spatial momenta as already demonstrated for mesons. These extensions will also allow for the calculation of transport coefficients like the shear viscosity. Other extensions that are left to future studies include the improvement of the presently used truncation by introducing wavefunction renormalization factors or a scale-dependent Yukawa coupling. Replacing the quarks here by nucleon fields and their parity partners allows to study the corresponding baryonic spectral functions in the parity-doublet model with fluctuations beyond mean-field as in \cite{Weyrich:2015hha} in order to describe the liquid-gas transition of nuclear matter and the chiral transition at high baryon density in a unified framework. This can then furthermore be extended to include vector and axialvector mesons along the lines of \cite{JungRenneckeTripoltEtAl2017}  and study their spectral changes in dense nuclear matter.

\acknowledgments
This work was supported by the Deutsche Forschungsgemeinschaft (DFG) through
the grant CRC-TR 211 ``Strong-interaction matter under extreme conditions,''
the German Federal Ministry of Education and Research (BMBF) through grant 05P16RDFC1, and by the Helmholtz International Center for FAIR within the LOEWE program of the State of Hesse.
 
\appendix

\section{Definitions and Flow Equations}\label{app:defs}

The three-dimensional bosonic and fermionic regulator functions are given by
\begin{align}
R^B_{k}&=(k^2-\vec q^{\,2})\theta(k^2-\vec q^{\,2})\,,\\
R^F_{k}&=i \slashed{\vec q} (\sqrt{k^2/\vec q^{\,2}}-1)
\theta(k^2-\vec q^{\,2})\,.
\end{align}

The threshold functions appearing in Eq.~(\ref{eq:flow_pot}) are given by
\begin{align}
I_{\sigma,\pi}^{(1)} =
\frac{k^4}{6\pi^2}
\frac{1}{E_{\sigma,\pi}},
\qquad
I_\psi^{(1)} =
\frac{k^4}{3\pi^2}
\frac{1}{E_\psi},
\end{align}
where the effective quasi-particle energies read
\begin{align}
\label{eq:energies}
E_\alpha=\sqrt{k^2+m_\alpha^2}, \qquad \alpha \in \{\pi,\sigma,\psi\}\,,
\end{align}
and the effective meson masses have already been introduced in Eq.~(\ref{eq:masses}). The three-point vertex functions appearing in Eq.~(\ref{eq:flow_gamma2}) are given by
\begin{align}
\Gamma^{(3)}_{\bar \psi \psi  \phi_i}=h \begin{cases}1 &\text{for}\, i=0\\
\I\gamma^5\tau^i &\text{for}\, i=1,2,3 \end{cases}\,.
\end{align}

The generalized loop functions used in Eq.~(\ref{eq:flow_eq_2PF}) are defined as
\begin{align}
\mathcal{J}^{(X)}_{k,\alpha\beta}(\omega)&=\int\frac{d^3q}{(2\pi)^3}\,
J^{(X)}_{k,\alpha\beta}(\omega),
\end{align}
with $\alpha,\beta\in \{\sigma,\pi,\psi\}$ and $X\in \{A,B,C\}$. The loop function $J^{(A)}_{k,\alpha\beta}(\omega)$ is zero for vanishing external spatial momentum, $|\vec{p}|=0$. $J^{(B)}_{k,\alpha\beta}(\omega)$ and $J^{(C)}_{k,\alpha\beta}(\omega)$ are given by
\begin{align}
\label{eq:JB}
J^{(B)}_{k,\alpha\beta}(\omega)=
&-\frac{1}{(\omega+ i\epsilon+E_\alpha+E_\beta)}
\frac{\pm kh^2m_\psi}{4E_\alpha^3E_\beta}
\nonumber\\
&-\frac{1}{(\omega+ i\epsilon+E_\alpha+E_\beta)^2}
\frac{\pm kh^2m_\psi}{4E_\alpha^2E_\beta}
\nonumber\\
&+\frac{1}{(\omega+ i\epsilon-E_\alpha-E_\beta)}
\frac{\pm kh^2m_\psi}{4E_\alpha^3E_\beta}
\nonumber\\
&-\frac{1}{(\omega+ i\epsilon-E_\alpha-E_\beta)^2}
\frac{\pm kh^2m_\psi}{4E_\alpha^2E_\beta},
\end{align}
\begin{align}
\label{eq:JC1}
J^{(C)}_{k,\alpha\psi}(\omega)=
&-\frac{1}{(\omega+ i\epsilon+E_\alpha+E_\psi)}
\frac{ kh^2}{4E_\alpha^3}
\nonumber\\
&-\frac{1}{(\omega+ i\epsilon+E_\alpha+E_\psi)^2}
\frac{ kh^2}{4E_\alpha^2}
\nonumber\\
&-\frac{1}{(\omega+ i\epsilon-E_\alpha-E_\psi)}
\frac{ kh^2}{4E_\alpha^3}
\nonumber\\
&+\frac{1}{(\omega+ i\epsilon-E_\alpha-E_\psi)^2}
\frac{ kh^2}{4E_\alpha^2},
\end{align}
and
\begin{align}
\label{eq:JC2}
J^{(C)}_{k,\psi\alpha}(\omega)=
&-\frac{1}{(\omega+ i\epsilon+E_\psi+E_\alpha)^2}
\frac{ kh^2}{4E_\psi E_\alpha}
\nonumber\\
&+\frac{1}{(\omega+ i\epsilon-E_\psi-E_\alpha)^2}
\frac{ kh^2}{4E_\psi E_\alpha},
\end{align}
with
\begin{equation}
\pm=
\begin{cases}
+&\text{for}\,\, \alpha=\sigma \,\,\text{or}\,\, \beta=\sigma\\
-&\text{for}\,\, \alpha=\pi \,\,\text{or}\,\, \beta=\pi\end{cases}.
\end{equation}

These flow equations are very similar to the corresponding equations obtained in a one-loop calculation, see for example \cite{KitazawaKunihiroNemoto2014}. We note that the limit $\epsilon\rightarrow 0$ in the definition of the retarded two-point functions, Eq.~(\ref{eq:continuation}), can be performed analytically for the flow equation of the imaginary part of the two-point functions. This can be seen by rewriting the imaginary part of the loop functions by using the Dirac-Sokhotsky identities,
\begin{align}
\lim_{\epsilon \rightarrow 0}\text{Im}\frac{1}{\omega+ i\epsilon \pm E_\alpha\pm E_\beta}\rightarrow -\pi\delta(\omega\pm E_\alpha \pm E_\beta),\\
\lim_{\epsilon \rightarrow
	0}\text{Im}\frac{1}{(\omega+ i\epsilon \pm E_\alpha\pm E_\beta)^2}\rightarrow \pi\delta'(\omega \pm E_\alpha\pm E_\beta).
\end{align}
The flow equation for the imaginary part of the retarded two-point function then reduces to a sum over a few values $k_0$ that correspond to the scales where one of the arguments of the delta function becomes zero, see \cite{JungRenneckeTripoltEtAl2017} for details.

\bibliography{qcd}

\end{document}